\documentclass[letterpaper, 11pt]{article}
\usepackage[utf8]{inputenc}
\usepackage{cite,amsmath,bbm,amssymb,amsthm,xifthen,mathtools,enumerate,xspace}
\usepackage[colorinlistoftodos]{todonotes}
\usepackage[multiple]{footmisc}
\usepackage{braket,titlesec,algorithm}
\usepackage[noend]{algpseudocode}
\usepackage{hyperref} 
\usepackage[inner=2.5cm,outer=2.5cm,bottom=2.5cm]{geometry}  
\usepackage[capitalise]{cleveref}
\usepackage{enumitem}
\usepackage{suffix} 
\usepackage{bbm} 
\usepackage{environ} 
\setlist[itemize]{noitemsep}

\newcommand{\CONGEST}{\ensuremath{\mathsf{CONGEST}}\xspace}

\newcommand{\CC}{\ensuremath{\mathsf{CLIQUE}}\xspace}
\newcommand{\CLIQUE}{\ensuremath{\mathsf{CLIQUE}}\xspace}
\newcommand{\CCL}{\ensuremath{\mathsf{CLIQUE\ +\ Lenzen's\ Routing}}\xspace}

\newcommand{\bandwidth}{\ensuremath{\mathsf{Global}\mathsf{Congestion}}\xspace}   
\newcommand{\GlobalCongestion}{\ensuremath{\mathsf{Global}\mathsf{Congestion}}\xspace}   
\newcommand{\capacity}{\ensuremath{\mathsf{LocalCongestion}}\xspace} 
\newcommand{\LocalCongestion}{\ensuremath{\mathsf{LocalCongestion}}\xspace} 
\newcommand{\congestion}{\ensuremath{\mathsf{congestion}}\xspace}
\newcommand{\dilation}{\ensuremath{\mathsf{dilation}}\xspace}
\newcommand{\communicationDegreeComplexity}{\ensuremath{\mathsf{Communication\ Degree\ Complexity}}\xspace}

\newcommand\bigO[1]{\ensuremath{{O}(#1)}}
\WithSuffix\newcommand\bigO*[1]{\ensuremath{{O}\left(#1\right)}}
\newcommand\tildeBigO[1]{\ensuremath{{\tilde{O}}(#1)}}
\WithSuffix\newcommand\tildeBigO*[1]{\ensuremath{{\tilde{O}}\left(#1\right)}}
\newcommand\littleO[1]{\ensuremath{\operatorname{o}(#1)}}
\WithSuffix\newcommand\littleO*[1]{\ensuremath{{o}\left(#1\right)}}
\newcommand\bigOmega[1]{\ensuremath{{\Omega}(#1)}}
\WithSuffix\newcommand\bigOmega*[1]{\ensuremath{{\Omega}\left(#1\right)}}
\newcommand\littleOmega[1]{\ensuremath{{\omega}(#1)}}
\WithSuffix\newcommand\littleOmega*[1]{\ensuremath{{\omega}\left(#1\right)}}
\newcommand\E[1]{\ensuremath{\operatorname{E}[#1]}}
\WithSuffix\newcommand\E*[1]{\ensuremath{\operatorname{E}\left[#1\right]}}

\newcommand{\interval}[1]{\ensuremath{\left[#1\right]}}
\newcommand{\rightopeninterval}[1]{\ensuremath{\left[#1\right)}}

\newcommand{\Indicator}{\mathbbm{1}}

\DeclareMathOperator{\poly}{poly}
\DeclareMathOperator{\polylog}{poly\log}

\DeclarePairedDelimiter{\size}{\lvert}{\rvert}
\DeclarePairedDelimiter{\ceil}{\lceil}{\rceil}
\DeclarePairedDelimiter{\floor}{\lfloor}{\rfloor}

\makeatletter
\newtheorem*{rep@theorem}{\rep@title}
\newcommand{\newreptheorem}[2]{%
\newenvironment{rep#1}[1]{%
 \def\rep@title{#2 \ref{##1}}%
 \begin{rep@theorem}}%
 {\end{rep@theorem}}}
\makeatother

\usepackage{thmtools}
\usepackage{thm-restate}

\makeatletter
\def\namedlabel#1#2{\begingroup
    #2%
    \def\@currentlabel{#2}%
    \phantomsection\label{#1}\endgroup
}
\makeatother

\newcommand*{\fullrefsingle}[1]{\hyperref[{#1}]{\cref*{#1} \emph{(\nameref*{#1})}}} 

\makeatletter
\newcommand{\fullref}[1]{%
  \@tempswafalse
  \@for\next:=#1\do
    {\if@tempswa\ and \else\@tempswatrue\fi\fullrefsingle{\next}}%
}
\makeatother

\makeatletter
\renewenvironment{proof}[1][\proofname]{%
   \par\pushQED{\qed}\normalfont%
   \topsep6\p@\@plus6\p@\relax
   \trivlist\item[\hskip\labelsep\bfseries#1\@addpunct{.}]%
   \ignorespaces
}{%
   \popQED\endtrivlist\@endpefalse
}
\makeatother

\makeatletter
\newenvironment{proofof}[1][]
{%
 \def\firstargument{#1}
 \def\noargumentspecified{}
 \ifx\firstargument\noargumentspecified
  \begin{proof}
 \else
  \begin{proof}[Proof of \autoref{#1}]
  \label{proof:#1}
 \fi
 {}
} %
{\end{proof} }

\newtheorem{theorem}{Theorem}[section]
\newreptheorem{theorem}{Theorem}
\newtheorem{claim}[theorem]{Claim}
\crefname{claim}{Claim}{Claims}
\newtheorem{lemma}[theorem]{Lemma}
\crefname{lemma}{Lemma}{Lemmas}

\newtheorem{definition}[theorem]{Definition}

\makeatletter
\newcommand*{\whp}{%
    \@ifnextchar{.}%
        {w.h.p}%
        {w.h.p.\@\xspace}%
}
\makeatother

\makeatletter
\newcommand*{\Wlog}{%
    \@ifnextchar{.}%
        {w.l.o.g}%
        {w.l.o.g.\@\xspace}%
}
\makeatother

\NewEnviron{killcontents}{}

\title{Near-Optimal Scheduling in the Congested Clique }
\author{%
    Keren Censor-Hillel
    \and Yannic Maus
    \and Volodymyr Polosukhin
}
\date{Technion
\footnote{\{ckeren, yannic.maus, po\}@cs.technion.ac.il}}

\begin{document}
\thispagestyle{empty}

\maketitle
\setcounter{page}{1}

\begin{abstract}

This paper provides three nearly-optimal algorithms for scheduling $t$ jobs in the \CLIQUE model. 
First, we present a deterministic scheduling algorithm that runs in $O(\GlobalCongestion + \dilation)$ rounds for jobs that are sufficiently efficient in terms of their memory. The \dilation is the maximum round complexity of any of the given jobs, and the \GlobalCongestion is the total number of messages in all jobs divided by the per-round bandwidth of $n^2$ of the \CLIQUE model. Both are inherent lower bounds for any scheduling algorithm. 

Then, we present a randomized scheduling algorithm which runs $t$ jobs in $O(\GlobalCongestion + \dilation\cdot\log{n}+t)$ rounds and only requires that inputs and outputs do not exceed $O(n\log n)$ bits per node, which is met by, e.g., almost all graph problems.
Lastly, we adjust the \emph{random-delay-based} scheduling algorithm [Ghaffari, PODC'15] from the \CONGEST model and obtain an algorithm that schedules any $t$ jobs in $O(t / n + \LocalCongestion + \dilation\cdot\log{n})$ rounds, where the $\LocalCongestion$ relates to the congestion at a single node of the \CLIQUE.  We compare this algorithm to the previous approaches and show their benefit.

We schedule the set of jobs on-the-fly, without a priori knowledge of its parameters or the communication patterns of the jobs. In light of the inherent lower bounds, all of our algorithms are nearly-optimal.

We exemplify the power of our algorithms by analyzing the message complexity of the state-of-the-art MIS protocol [Ghaffari, Gouleakis, Konrad, Mitrovic and Rubinfeld, PODC'18], and we show that we can solve $t$ instances of MIS in $O(t + \log\log\Delta\log{n})$ rounds, that is, in $O(1)$ amortized time, for $t\geq \log\log\Delta\log{n}$.

\end{abstract}
\section{Introduction}
\label{sec:intro}
Motivated by the ever-growing number of frameworks for parallel computations, we address the complexity of executing multiple jobs in such settings. Such frameworks, e.g., MapReduce \cite{Karloff2010}, typically need to execute a long queue of jobs. 
A fundamental goal of such systems is to schedule many jobs in parallel, for utilizing as much of the computational power of the system as possible. Ideally, this is done by the system in a black-box manner, without the need to modify the jobs and, more importantly, without the need to know their properties  and specifically their communication patterns beforehand.

In their seminal work, Leighton, Maggs, and Rao \cite{LeightonMR94} studied the special case where each of the to-be-scheduled jobs is a  routing protocol that routes a packet through a network along a given path. The goal in their work is to schedule $t$ jobs such that the \emph{length} of the schedule, i.e., the overall runtime until all $t$ packets have reached their destination, is minimized.
They showed that there exists an optimal packet-routing schedule of length $\bigO{\congestion + \dilation}$, where  \congestion is the maximum number of packets that need to be routed over a single edge of the network and \dilation is the maximum length of a path that a packet needs to travel. Clearly, both parameters are lower bounds on the length of any schedule, implying that the above schedule is asymptotically optimal. Further, Leighton, Maggs, and Rao  \cite{LeightonMR94} showed that assigning a random delay to each packet gives a schedule of length $\bigO{\congestion + \dilation\cdot \log{(t\cdot \dilation )}}$.

In his beautiful work, Ghaffari \cite{Ghaffari2015} raised the question of running multiple jobs in the distributed \CONGEST model on $n$ nodes. Applying the random delays method \cite{LeightonMR94}, he showed a randomized algorithm which after $\bigO{\dilation\cdot\log^2{n}}$ rounds of pre-computation, runs a given a set of jobs in $\bigO{\congestion + \dilation\cdot\log{n}}$ rounds. Here, in a similar spirit to \cite{LeightonMR94},  \congestion is the maximum number of messages that need to be sent over a single edge and \dilation is the maximum round complexity of all jobs.
Further, Ghaffari \cite{Ghaffari2015} showed that this is nearly optimal, by constructing an  instance which requires $\Omega{(\congestion + \dilation\cdot\log{n}/\log\log{n})}$ rounds to schedule.

In this paper, we address the $t$-scheduling problem in the ($\mathsf{CONGESTED}$) \CC model \cite{LotkerPPP05}, in which each of $n$ machines can send $O(\log n)$-bit messages to any other machine in each round.
Our goal is thus to devise scheduling algorithms that run $t$ jobs in a black-box manner, such that they complete in a number of rounds that beats the trivial solution of simply running the jobs sequentially one after the other, and, ideally, reaches inherent lower bounds that we discuss later.
We emphasize that we schedule all jobs' actions \emph{on-the-fly} during their execution. 
Throughout the paper, we use the terminology that a job is a \emph{protocol} that $n$ \emph{nodes}, $v_0,\dots, v_{n-1}$, need to run on some input, and we use the notion of an \emph{algorithm} for the scheduling procedure that the $n$ \emph{machines}, $p_0,\dots, p_{n-1}$, execute. Each machine $p_i$ is given the inputs of the nodes $v^{j}_i$ for all jobs $j$, and the machines run an algorithm which simulates the protocols of their assigned nodes.

Our contributions are three algorithms for scheduling $t$ jobs in the \CC model, which exhibit trade-offs based on the parameters of \dilation, \capacity, and \bandwidth of the set of jobs, which we formally define below. Our scheduling algorithms complete within round complexities that are nearly optimal w.r.t. the appropriate parameters. 

\subsection{Our Contributions}

No scheduling algorithm can beat the \dilation of the set of jobs, which is the maximum runtime of a job in the set, had this job been executed standalone. Similarly, another natural lower bound is given by the \bandwidth, which is the total number of messages that all nodes in all jobs send over all rounds, normalized by the $n^2$ per-round-bandwidth of the \CC model (for simplicity, this considers the possibility that a machine sends a message to itself).
The main goal is thus to get as close as possible to these parameters.

As a toy example, consider a set of jobs in which each completes within a single round. Intuitively, if the total number of messages that need to be sent by all nodes in all jobs is at most $n^2$, then one could hope to squeeze all of these jobs into a single round of the \CC model, as $n^2$ is the available bandwidth per round. The main hurdle in a straightforward argument as above, lies in the fact that a machine cannot send more than $n$ messages in a round. Thus, although we are promised that in total there no more than $n^2$ messages, it might be that a machine is required to send/receive $\omega(n)$ messages because the heaviest-loaded nodes of multiple jobs might be located on the same machine. 

This implies that a na\"ive scheduling, in which each machine simulates the nodes that are located at it, is more expensive than our single-round goal scheduling, as some messages must wait for later rounds. In the general case, these issues become more severe, as the jobs may originally require more than a single round, and it could be that each round displays an imbalance in a different set of nodes and machines.

The key ingredient in the first two scheduling algorithms that we present is hence to \emph{rebalance} the nodes among the machines, for the sake of a more efficient simulation that deals with the possible imbalance, which also may vary from round to round. 
The third scheduling algorithm we present is inspired by the random-delay approach of \cite{LeightonMR94,Ghaffari2015}. In what follows, we present the guarantees that are obtained by our three scheduling algorithms, and discuss the trade-offs that they exhibit.

\paragraph{Deterministic scheduling.}  
A crucial factor in the complexity of rebalancing the nodes among the machines is the amount of information that needs to be passed from one machine to another in order for the latter to take over the simulation of a node.
To this end, we define an $M$-\emph{memory efficient} job as a job where for each node, its state can be encoded in $M\log{n}$ bits, and that the number of messages it needs to receive in this round can be inferred from its state.
In \cref{sec:deterministic}, we obtain the following deterministic algorithm for scheduling $t$ jobs that are $M$-memory efficient.

\begin{restatable*}
{theorem}{deterministic}\label{thr:optimalShedulingOfSmallJobs}
There is a deterministic algorithm that schedules $t=\poly{n}$ jobs that are $M$-memory efficient 
 in $\bigO{\bandwidth+\ceil{M\cdot t/n}\cdot\dilation}$ rounds.
\end{restatable*}

At a very high level, in the algorithm for \cref{thr:optimalShedulingOfSmallJobs}, the machines rebalance nodes in each round by sending the states of nodes. The main technical effort is that the reassignment needs to be computed by the machines \emph{on-the-fly}, and we show how to do so in a fast way.

Notice that for the case that $M\cdot t=\bigO{n}$, the round complexity we get from \cref{thr:optimalShedulingOfSmallJobs} is $O(\bandwidth+\dilation)$, which is \emph{optimal}.
Another crucial point is that our algorithm does not require the knowledge of either the \bandwidth or the \dilation of the set of jobs.

\paragraph{Randomized scheduling.}
If we are given a set of jobs that are not memory efficient for a reasonable value of $M$,  it may be too expensive to rebalance the nodes among the machines in every simulated round. However, if the input of each node is not too large, we can randomly shuffle the nodes at the beginning of the simulation, and if the output is also not too large then we can efficiently unshuffle, and reach the original assignment.

To capture this, we say that a job is \emph{I/O efficient} if its input and output can be encoded within $\bigO{n\log{n}}$ bits. Notice that most graph-related problems are I/O efficient, e.g., MST \cite{LotkerPPP05,HegemanPPSS15,GhaffariP16,Korhonen16,Jurdzinski018,Nowicki19}, MIS \cite{Ghaffari17,GhaffariGKMR18,Censor-HillelPS20}, Mininum Cut \cite{GhaffariN18,Ghaffari0T20}, as well as many algebraic problems \cite{Censor-HillelKK19,Gall16}.
An example of a graph problem that is not I/O efficient is  \emph{$k$-clique listing}, in which all nodes together have to explicitly output \emph{all} $k$-cliques in the input graph \cite{DolevLP12,IzumiG17,Pandurangan0S18,Censor-HillelGL20,ChangPZ19}
which can be as many as $\Omega(n^k)$,
thus necessitating large outputs. While the $k$-clique listing problem is not \emph{output efficient}, it is \emph{input efficient}, and as it does not require a specific node to output a specific clique, one could also run several instances of the problem by omitting the output unshuffling step of our scheduling algorithm.

We obtain the following randomized algorithm for scheduling $t$ jobs that are I/O efficient.
\begin{restatable*}
{theorem}{shuffle}\label{thr:shuffle}
There is a randomized algorithm in the \CC model that schedules $t={\poly{n}}$ jobs that are I/O efficient  in
$\bigO{t + \bandwidth + \dilation\cdot \log{n}}$ 
rounds, \whp \footnote{\label{fn:whp}An event occurs \emph{\whp} (\emph{with high probability}) if for an arbitrary constant $c\geq 1$, the probability that the event occurs is at least $1 - n^{-c}$, where $n$ is the number of machines. All our results can be adapted to any constant $c$ at the cost of increasing the runtime by a constant factor.}
\end{restatable*}

As the deterministic scheduling algorithm (\cref{thr:optimalShedulingOfSmallJobs}), the scheduling algorithm of \cref{thr:shuffle} requires neither the knowledge of \bandwidth nor the knowledge  of  \dilation.

Both of our scheduling algorithms for \cref{thr:optimalShedulingOfSmallJobs} and \cref{thr:shuffle} have the machines possibly simulate the execution of nodes that are not originally assigned to them.
We stress that any black-box scheduling algorithm in which each machine only simulates the nodes that are originally assigned to it must inherently suffer from another type of congestion as a lower bound on its round complexity, namely, the maximum number of messages that all nodes assigned to a single machine have to send or receive, normalized by the bandwidth $n$ that each machine has per round. We call this the \capacity of a set of jobs. We obtain the following \emph{random-delay-based} algorithm for scheduling any $t$ jobs, without reassigning nodes.

\begin{reptheorem}{thr:delay}[Simplified]There is a randomized algorithm in the \CC model that
schedules $t={\poly{n}}$ 
jobs  in
$\bigO{{t/n} + \capacity + \dilation\cdot\log{n}}$ rounds \whp
\end{reptheorem}
The stated complexity in the above simplified version of \cref{thr:delay} requires the knowledge of the \capacity, but this can be eliminated using a standard doubling approach, at the cost of a logarithmic multiplicative factor (see precise statement in \cref{sec:randomized}). 

The random-delay algorithm which gives \cref{thr:delay} is suboptimal for a set of jobs which have a single machine with heavily-loaded nodes assigned to it, since in this case it does not exploit the entire bandwidth of the \CC model. For example, for a problem with inputs of at most $O(n\log n)$ bits per node, a protocol in which a fixed leader learns the entire input takes $\bigO{n}$ rounds, where on each round each node sends one message to the leader, who receives $n$ messages. For $n$ such jobs, the $\bandwidth$ is $n$, while the $\capacity$ is $n^2$. In such a setting,  our random-shuffling algorithm from \cref{thr:shuffle} outperforms the random-delay algorithm from \cref{thr:delay}. One may suggest to replace the fixed leader by a randomly or more carefully chosen leader. However, this trick might be more complicated in the general case: suppose now that $n^{0.9}$ nodes need to learn $n^{1.1}$ messages each. For such a set of jobs, it holds that  $\capacity=n^{0.1}$, while $\bandwidth=1$. Thus, it is more efficient to run \cref{thr:shuffle} in this case. Another crucial example in which random-shuffling outperforms random-delays is the maximal independent set protocol that we describe below. Note that our algorithms address these cases in a black-box manner without assuming knowledge of the communication pattern.

\paragraph{Applications.} In \cref{sec:applications},  we present two applications in order to exemplify our scheduling algorithms. 
We summarize these applications below and defer a more detailed discussion to \cref{sec:applications} and \cref{sec:discussion}.

A \emph{maximal independent set} (MIS) of a graph $G=(V,E)$ is a set $M\subseteq V$ such that no two nodes in $M$ are adjacent and no node of $V$ can be added to $M$ without violating this condition. The state-of-the-art randomized \CC protocol for solving the MIS problem completes in $O(\log \log \Delta)$ rounds, \whp, where $\Delta$ is the maximum degree of the graph \cite{GhaffariGKMR18}. We analyze the  message complexity of this protocol, and show that it does not utilize the entire bandwidth. Thus, we can schedule multiple MIS jobs efficiently using our random shuffling scheduling algorithm from \cref{thr:shuffle}, and we obtain the following theorem.

\begin{restatable*}[Multiple MIS instances]{theorem}{MISAmortized}\label{thr:MISAmortized}
There is a randomized algorithm in the \CC model which solves $t={\poly{n}}$ instances of MIS in $\bigO{t+\log\log\Delta\log n}$ rounds, \whp
\end{restatable*}

Another application that exemplifies our scheduling algorithms is a variant of the \emph{pointer jumping} problem, which is a widespread algorithmic technique \cite{Hirschberg76}. In the $P$-pointer jumping problem, each node has a permutation on $P$ elements. A fixed node has a value $pointer$ $p$ and should learn the result of applying these permutations one after another on $p$. Pointer jumping can be solved by an $O(\log{n})$-round protocol in the \CC model by learning the composition of all permutations (see \cref{subsec:pointerJumping}). We observe that this protocol does not utilize the entire bandwidth and leverage this for obtaining an algorithm that executes multiple instances of this protocol efficiently.

\begin{restatable*}[Pointer Jumping]{theorem}{PJAmortized}\label{thr:PJAmortized}
    For $P \leq n$, there are algorithms in the \CC model that solve $t={\poly{n}}$ instances of the $P$-pointer jumping problem  
    deterministically in $\bigO{\ceil{P\cdot t/n} \cdot \log{n}}$, and 
    randomized in $\bigO{t + \log^2{n}}$ rounds, \whp.
\end{restatable*}

We obtain the deterministic result using our scheduling algorithm in \cref{thr:optimalShedulingOfSmallJobs} and the randomized result using our random-shuffling scheduling algorithm in \cref{thr:shuffle}. The proposed simple $O(\log n)$ round pointer jumping protocol also serves as an example where scheduling jobs via the random-shuffling approach of \cref{thr:shuffle} is significantly better than the random-delay based approach of \cref{thr:delay}. For more details we refer to \cref{subsec:pointerJumping}.

\medskip

In \cref{sec:discussion} we discuss the amortized versions of these results, and present a small example of a set of jobs that can be scheduled with $o(1)$-amortized complexity. In light of the growing number of $O(1)$-round \CC-protocols, e.g.,  \cite{CzumajDP20,Nowicki19,Ghaffari0T20}, we propose 
the amortized complexity of solving many instances of a problem in parallel, as a valuable measure for the efficiency in future research.

\subsection{Related Work}
Many graph problems are studied in the \CC model. 
There are fast protocols for the \CC model for
distance computations \cite{Censor-HillelKK19,Gall16},
minimum spanning tree (MST) \cite{LotkerPPP05,GhaffariP16,Korhonen16,Nowicki19}, MIS \cite{Ghaffari17,GhaffariGKMR18,Censor-HillelPS20},
and more.

To the best of our knowledge, there are no previous works that study the scheduling of jobs in the \CC model. In the past, it has been shown that running multiple instances of \emph{the same} protocol on different inputs can result in fast algorithms for some complex problems. We survey some of these.
Hegeman et al.\ \cite{HegemanPPSS15} 
reduce the MST problem to multiple smaller instances of graph connectivity, breaking below the long-standing upper bound of $O(\log\log{n})$ by Lotker et al.\  \cite{LotkerPPP05}. Further variants and improvements on the MST problem \cite{Korhonen16,GhaffariP16,Jurdzinski018,Nowicki19} all exploit invoking multiple instances of sparser problems. This line of work culminated in the  deterministic $\bigO{1}$-round algorithm of Nowicki \cite{Nowicki19}.

In \cite{GhaffariN18}, Ghaffari and Nowicki show a randomized algorithm which solves $\bigO{n^{1-\epsilon}}$ many instances of the MST problem in $\bigO{\epsilon^{-1}}$ rounds. This is used for finding the minimum cut of a graph. The state-of-the-art $\bigO{1}$-round algorithm for the minimum cut problem, by Ghaffari et al.\  \cite{Ghaffari0T20}, runs $\Theta{(\log{n})}$ instances of connected components as a subroutine.
The complexity of computing multiple matrix multiplications in parallel was explored by Le Gall \cite{Gall16} and was used in the same paper to solve the \emph{all-pairs-shortest-path problem}.

The notion of \capacity is somewhat similar to the notion of \communicationDegreeComplexity \cite{KlauckNP015}. The difference lies in the fact that the  \communicationDegreeComplexity is an upper bound on the number of messages sent or received by any node on \emph{any round}, while \capacity is an upper bound on the \emph{total} number of messages sent or received by any node \emph{over all} rounds. 

\section{Preliminaries}
\label{sec:model}
\paragraph{The \CC Model.}
In the ($\mathsf{CONGESTED}$) \CC model, $n$ machines $p_0,\ldots,p_{n-1}$  communicate with each other in synchronous rounds in an all-to-all fashion. In each round, any pair of machines can exchange $\bigO{\log{n}}$ bits. There is usually no constraint neither on the size of the \emph{local memory} nor on the time complexity of the \emph{local computations}. Besides the local memory, each machine has a \emph{read-only input buffer} and a \emph{write-only output buffer}, as well as \emph{read/write incoming- and outgoing- message buffers}. 

\vspace{-0.3cm}
\paragraph{Routing in the \CC Model.}
Lenzen's routing scheme \cite{Lenzen13} says that a set of messages can be routed in the \CC model within $O(1)$ rounds, given that each machine sends and receives at most $O(n)$ messages. We formally state it here in its generalized version, which addresses the case of more than a linear number of messages. In the generalized version,  each machine $p_i$ holds a set of messages $M_i=\bigcup_{i'\in [n]}{M_i^{i'}}$, where $M_i^{i'}$ is a set of messages with the destination $p_{i'}$. The claim follows by having each node chop its set of messages $M_i$ into chunks of $n$ messages, each of which containing $|M_i^{i'}| n/X$ messages for each $i'\in [n]$, and applying the original routing scheme $X/n$ times. The routing scheme could be adapted to preserve the message complexity in the following way.\footnote{We thank an anonymous reviewer for pointing this out.} Let $Y=\sum_{i\in\interval{n}}\size{M_i}\leq n^2$ be the total number of messages. First, compute a global numbering of messages and the total number of messages $Y$. Then, send $\bigO{\ceil{\sqrt{Y}}}$ messages to each one of the first $\bigO{\ceil{\sqrt{Y}}}$ machines via intermediate nodes based on the numbering. Sort messages by the destination in the using Lenzen's sorting algorithm \cite{Lenzen13} over $\bigO{\ceil{\sqrt{Y}}}$-clique. Finally, deliver the messages to their destinations via intermediate nodes based on the indices of messages in the sorted sequence. The round complexity of the algorithm is $\bigO{1}$ and the message complexity of the algorithm in $\bigO{Y+\ceil{\sqrt{Y}}\cdot\ceil{\sqrt{Y}}}=\bigO{Y}$

\begin{claim}[Lenzen's Routing Scheme]
\label{thr:optimalRouting}
Let $X$ be a globally known value and let $\mathcal{P}$ be the property that $|M_i | \leq X$ for all $i \in [n]$ and $\sum_{i\in [n]}{|M_i^{i'}|}\leq X$ for all $i' \in [n]$. 
There is an algorithm in the \CC model which completes in $O(\ceil{X/n})$ rounds and $O(\sum_{i\in\interval{n}}{M_i})$ messages, and delivers all messages if $\mathcal{P}$ holds, or indicates that it does not hold.
\end{claim}

\vspace{-0.4cm}
\paragraph{Protocols and Jobs.} 
A \emph{protocol} is run on an \emph{input}, that is provided in a distributed manner in the \emph{read-only input buffer} of each machine. The \emph{complexity} of a protocol is the number of synchronous rounds until each machine has finished writing its output to its \emph{write-only output buffer}.

A \emph{job} is an instance of a protocol together with a given input and a job is \emph{finished} when each machine has written its output. We generally assume that each job finishes in  $\bigO{\poly{n}}$ rounds.

For our purposes of fast scheduling, we need to specify the internals of each synchronous round. We follow the standard description, which is usually omitted and simply referred to as a 'round'. We require that for each machine, the input and output buffers are only accessed in the first and last rounds of the protocol on that machine,
respectively. In particular, this means that any further access to the input requires storing it in the local memory. Accessing the incoming- and outgoing-message buffers is not restricted to certain rounds. Each synchronous round of a protocol consists of $3$ steps, in the following order. 
\begin{itemize}[nosep]
    \item \textbf{\namedlabel{phase:reading}{Receiving Step}{\label{itm:readingInputMessages}:}} Read from incoming-message buffer (or from input buffer if this is the first round), possibly modifying the local memory.
    \item \textbf{\namedlabel{phase:computation}{Computation Step}{\label{itm:localComputationStep}:}} Possibly modify local memory.
    \item \textbf{\namedlabel{phase:writing}{Sending Step}{\label{itm:writingOutputMessages}:}} Write to outgoing-message buffer, (or to output buffer if this is the last round), possibly modifying the local memory.
\end{itemize}
After these 3 phases, all messages written in outgoing-message buffers are delivered into the incoming-message buffers of their targets.

\vspace{-0.3cm}
\paragraph{The Scheduling Problem.} 
In the \emph{$t$-scheduling problem} (or simply a scheduling problem, if $t$ is clear from the context) the objective is to execute $t$ jobs. Since our goal is to do this in an efficient manner, we wish to allow a machine to simulate a computation that originally should take place in a different machine, in a na\"ive execution of the $t$ jobs. To this end, we distinguish between the physical machine and the \emph{nodes}, which are the virtual machines that need to execute each job. That is, for each job $j$ we denote by $\{v_{i, j} | i \in [n]\}$ the set of nodes that need to execute job $j$. 

Formally, in the $t$-scheduling problem, the input for machine $p_i$ is composed of the inputs of all the nodes with identifiers of the form $v_{i, j}$ for each job $j\in[t]$. We also assume that each machine knows the protocol for each of the $t$ jobs. An algorithm 
\emph{solves the scheduling problem} or \emph{schedules the jobs} when each job has finished writing its output. That is, for deterministic jobs, we require each machine $p_i$ to write the output of nodes $v_{i, j}$ for all $j\in [t]$. For randomized jobs, the machines' output distribution for each job has to be equal to the distribution of outputs in a na\"ive execution of the job.  In the rest of the paper, we refer to the scheduling solution as an \emph{algorithm}, while we use the term \emph{protocol} only for the content of a job.

\vspace{-0.3cm}
\paragraph{Notations.}
Following the widespread conventions, we denote by $\log$ the logarithm base $2$, and by $\ln$ the natural logarithm. Also, we denote $\interval{n}=\set{0, 1, \ldots, n - 1}$.
We denote by $s_{i, j}^r$ and $t_{i, j}^r$ the number of messages sent and received by $v_{i, j}$ in round $r$, respectively.  If job $j$ terminates before round $r$, we indicate $s_{i, j}^r=t_{i, j}^r=0$. We sometimes drop the superscript $r$, when it is clear from the context. We denote by $\ell_j$ the round complexity of  job $j$ and by 
$m_j
=\sum_{i\in\interval{n},r\in\interval{\ell_j}}{s_{i, j}^r}
=\sum_{i\in\interval{n},r\in\interval{\ell_j}}{t_{i, j}^r}$
the total number of messages sent or received during the execution of job $j$, i.e., the message complexity of  job $j$. Another notation we extensively use is 
$m^r
=\sum_{i\in\interval{n},j\in\interval{t}}{s_{i, j}^r}
=\sum_{i\in\interval{n},j\in\interval{t}}{t_{i, j}^r}$
, which is the number of messages all nodes in all jobs sent or received during  round $r$.

\vspace{-0.3cm}
\paragraph{Congestion parameters.} We define the normalized \bandwidth as the total number of messages sent by all the jobs divided by $n^2$, and normalized \capacity as the maximum number of messages send to or received by some node in the entire course of the execution of all jobs divided by $n$. Formally, $\dilation
    =\max_{j\in\interval{t}}{\ell_j}$,
\begin{align*}
    \bandwidth&
    =\sum_{j\in\interval{t}}{m_j}
    ={\sum_{i\in\interval{n}}\sum_{j\in\interval{t}}\sum_{r\in\interval{\ell_j}}{s_{i, j}^r}}/{n^2}
    =\sum_{r\in\interval{\dilation}}{m^r}/{n^2},
    \\
    \capacity&
    =\max\Set{
\max_{i\in\interval{n}}{\sum_{j\in\interval{t}}\sum_{r\in\interval{\ell_j}} s_{i, j}^r} / n, 
\max_{i\in\interval{n}}{\sum_{j\in\interval{t}}\sum_{r\in\interval{\ell_j}} t_{i, j}^r} / n}.
\end{align*}

\paragraph{Hoeffding bound.}
Some of our proofs use the following Hoeffding bound.
\begin{claim}[Hoeffding Bound \cite{Hoeffding1963}] 
Let $\Set{X_i}_{i=1}^{n}$ be independent random variables with values in the interval $X_i\in\interval{0, 1}$ and expectation of their sum bounded by $E\left[\sum_{i=1}^{n}{X_i}\right]\leq\mu$. Then for all $\epsilon>0$
\begin{align*}
        \Pr\left[
            \sum_{i=1}^{n}{X_i}\geq \left(1+\epsilon\right)\mu
        \right]
        \leq
        \left(
            \frac{e^\epsilon}{\left(1+\epsilon\right)^{1+\epsilon}}
        \right)^{\mu}
        \leq 
        e^{-\frac{\epsilon^2}{2+\epsilon}\mu}~.
    \end{align*}
\label{thr:relativeHoeffding}
\end{claim}

\section{Deterministic Scheduling}\label{sec:deterministic}
The objective of this section is to prove the following theorem.
\deterministic

The formal definition of an $M$-memory efficient job  as used in \cref{thr:optimalShedulingOfSmallJobs} is as follows.
\begin{definition}[\textbf{$M$-memory efficient job}] For a given value $M$, an \emph{$M$-memory efficient job} is a job in which for each node $v$ in each round $r$, the state (local memory) of $v$ at the end of the \ref{phase:computation} can be encoded in $M\log{n}$ bits. In addition,
there is a function that, given the state of node $v$ after the \ref{phase:computation} of round $r$, infers the number of messages it sends and receives on this round. 
\end{definition}

\Cref{thr:optimalShedulingOfSmallJobs} requires that jobs use at most $M$ bits of local memory per machine. Thus, the power of the result is when $M=\littleO{n}$, as otherwise the na\"ive execution of jobs one after another 
schedules them in $\dilation\cdot t$ rounds.  In the case that $M\cdot t=\bigO{n}$, the runtime becomes $O(\bandwidth + \dilation)$, which is optimal up to a constant factor as, clearly, any schedule  for any collection of jobs
requires at least $\bigOmega{\bandwidth + \dilation}$ rounds.

 To schedule the jobs for \Cref{thr:optimalShedulingOfSmallJobs},  we work in epochs. Each machine $p_i$ first simulates round 0 up to the end of the \ref{phase:computation} for the nodes $v_{i,j}$, for each $j \in [t]$. This does not require any communication. Then, the epochs are such that for each round $r$, at the start of epoch $r$, all nodes in all jobs are at  the end of the \ref{phase:computation} of round $r$. 
 Clearly, for each simulated node that finishes in round $r$, the machine does not need to do anything for the part that executes the beginning of round $r+1$. The reason why we execute the protocol in these \emph{shifted} epochs, from \ref{phase:writing} of round $r$ (including) to \ref{phase:writing} of round $r+1$ (excluding), lies in the fact that the bottleneck is the possible  imbalance in communication. 
 
Recall that  $m^r$ denotes the number of messages all nodes from all jobs send in round $r$. Since in each round of the \CC model, at most $n^2$ messages can be exchanged, routing $m^r$ messages cannot be done faster than $\ceil{{m^r}/{n^2}}$ rounds. We aim to execute an epoch in this optimal number of $\bigO{\ceil{{m^r}/{n^2}}}$ rounds. We start with the simple case and then use it to solve the general case. 

The first case is when $m^r\leq  2n^2$. In \cref{thr:n2messagesFromSqrtNJobsIn1Round}, we show that in this case, we can route all $m^r$ messages in $O(\ceil{M\cdot t/n})$ rounds.  The challenge we encounter is that although $m^r\leq 2n^2$, we are not promised that the messages are balanced across the machines in the following sense. It is possible that some machine $p_i$, which simulates the nodes $v_{i, j}$, for all jobs $0\leq j < t$, is required to send significantly more than $n$ messages when summing over all messages that need to be sent by these nodes $v_{i, j}$. We overcome this issue by assigning the simulation of some of these nodes to some other machine $p_{i'}$, which originally has a smaller load of messages to send. The crux that underlies our ability to defer a simulation of a node $v_{i,j}$ to a machine $p_{i'}$ is that the state of the node does not consume too many bits. We show how to compute a well-balanced assignment of nodes to machines in \cref{thr:reassigningNodeRolesInJobs}. This assignment allows us to execute the epoch in the claimed number of $O(\ceil{M\cdot t / n})$ rounds. 

In the general case, we can have $m^r > 2n^2$.
We show how to carefully split up the messages that need to be sent into chunks that allow us to use multiple invocations of \cref{thr:n2messagesFromSqrtNJobsIn1Round}. This allows us to execute the epoch in the $\bigO{\ceil{{m^r}/{n^2} + M\cdot t / n}}$ rounds.
As the core of our algorithm is handling the case $m^r\leq 2n^2$, now, we focus on the case $m^r\leq 2n^2$.

We start with the following notation.
An \emph{assignment} of nodes to machines corresponds to a function $\varphi\colon\interval{n}\times\interval{t}\mapsto\interval{n}$, where $\varphi(i,j)=k$ says that the $i$-th node in job $j$, i.e.,  $v_{i,j}$, is assigned to the $k$-th machine $p_k$. We sometimes abuse notation and write that $\varphi(v_{i,j})=p_k$ for $\varphi(i,j)=k$. We call an assignment \emph{balanced}, if the number of nodes assigned to each machine is $\bigO{t}$, i.e., if for each $k$, it holds that $\size{\varphi^{-1}(p_k)}=\bigO{t}$. The (balanced) assignment $\varphi(i, j)=i$ is called the \emph{trivial} assignment.

We denote by $S_{i, j, r}$  the state of node $v_{i,j}$ after its \ref{phase:computation} in round $r$.
\begin{restatable}[Distributing the states]{claim}{reassigningNodeRolesInJobs}\label{thr:reassigningNodeRolesInJobs}
    Given are $t$ jobs that are $M$-memory efficient, and globally known initial and final balanced assignments, $\varphi_s$ and $\varphi_f$, respectively. Assume that for each $i\in [n]$ and $j\in [t]$, machine $\varphi_s(i,j)$ holds the state $S_{i, j, r}$ of node $v_{i,j}$ after its \ref{phase:computation} in round $r$. 
	Then, there exists a deterministic \CC algorithm which completes in $\bigO{\ceil{M\cdot t/n}}$ rounds and \emph{moves} the states according to $\varphi_f$, that is, at the end of the algorithm, for each $i\in [n]$ and $j\in [t]$, machine $\varphi_f(i,j)$ holds the state  $S_{i, j, r}$ of node $v_{i,j}$.
\end{restatable}
\begin{proofof}[thr:reassigningNodeRolesInJobs]
For each node $v_{i, j}$, denote $i'=\varphi_f(i, j)$.
For each node $v_{i, j}$ such that $i''=\varphi_s(i, j)$, machine $p_{i''}$ sends $S_{i,j,r}$ to machine $p_{i'}$.
Overall, each machine $p_i$ sends and receives $\size{\varphi_s^{-1}(p_i)}\cdot M=\bigO{t\cdot M}$, $\size{\varphi_f^{-1}(p_i)}\cdot M=\bigO{t\cdot M}$ messages.
Thus, by \cref{thr:optimalRouting}, it completes in $\bigO{\ceil{{M\cdot t}/{n}}}$ rounds. 
\end{proofof}
\begin{lemma}[Scheduling of a round with $m^r\leq 2 n^2$ messages]\label{thr:n2messagesFromSqrtNJobsIn1Round}
Given are $t$ jobs that are $M$-memory efficient, and given is a round number, $r$, for which $m^r\leq 2n^2$. Assume that for each $i \in [n]$, $p_i$ holds $S_{i,j,r}$ for all $j\in [t]$. Then there exists a deterministic \CC algorithm which completes in $\bigO{\ceil{M\cdot t/n}}$ rounds, at the end of which, for each $i \in [n]$, $p_i$ holds $S_{i,j,r+1}$ for all $j\in [t]$.
\end{lemma}

The outline of the algorithm is as follows. Each machine partitions its simulated nodes into buckets of contiguous ranges of indices, such that nodes in each bucket send and receive $\bigO{n}$ messages altogether. Thus, the messages of all nodes in the bucket can be sent or received by a single machine.
We show that the number of buckets over all machines is $\bigO{n}$. The machines collectively assign the buckets such that each machine gets $\bigO{1}$ buckets, and they make the assignment globally known.
Then, the states $S_{i,j,r}$ are distributed according to the assignment using \cref{thr:reassigningNodeRolesInJobs}, and each machine executes the \ref{phase:writing} of round $r$ for each of its newly assigned nodes and all messages get delivered. Then, each machine executes the remainder of the protocol of its newly assigned nodes until after the \ref{phase:computation} of round $r+1$. Finally, the states $S_{i,j,r+1}$ for round $r+1$ are distributed back according to the trivial assignment.

\begin{proofof}[thr:n2messagesFromSqrtNJobsIn1Round]

We begin with describing the algorithm (see \cref{alg:atMostN2}). Afterwards, we prove the correctness and analyze the round complexity. 

\begin{algorithm}
\caption{Simulating a round with $m^r\leq n^2$.}
\label{alg:atMostN2}
\begin{algorithmic}[1]
    \State{\label{step:atMostN2:assign}Compute the balanced     assignment $\varphi\colon\interval{n}\times\interval{t}\mapsto\interval{n}$.    }
    \State{\label{step:atMostN2:move}Distribute the states according to the assignment $\varphi$.}
    \State{\label{step:atMostN2:execute}Execute the protocol for round $r$ accounting for $\varphi$.}
    \State{\label{step:atMostN2:return}Distribute the states back according to the trivial assignment.}
\end{algorithmic}
\end{algorithm}

\textbf{The Algorithm.}
We first show how to split nodes into buckets. Then we show how to compute a globally known assignment $\varphi$, distribute the nodes according to $\varphi$, execute the jobs until after the next \ref{phase:computation}, and assign nodes back to their initial machines. 

\textit{Forming buckets (locally):} 
Each machine $p_i$ for each $j\in[t]$ uses $S_{i,j,r}$ to locally compute $s_{i, j}$ and $t_{i, j}$, the number of messages each node $v_{i, j}$ sends and receives in round $r$, respectively. 
This is possible by the definition of an $M$-memory efficient job.
Let $S_i=\sum_{j=0}^{t-1}{s_{i, j}}$ and $T_i=\sum_{j=0}^{t-1}{t_{i, j}}$. Then, each machine $p_i$ (locally and independently) applies \cite[Lemma 7]{Censor-HillelDK19} (restated in \cref{thr:split} for better readability) with $k=k_i=\ceil{\max\set{S_i / n, T_i / n}}$ to the sequences $(s_{i, j})_{j=0}^{t-1}$ and $(t_{i, j})_{j=0}^{t - 1}$, to split its nodes into $k_i$ buckets $B_{i,0},\ldots,B_{i,k_i-1}$ of continuous ranges of jobs' indices. 

\begin{claim}[Lemma 7 from \protect{\cite{Censor-HillelDK19}}]\label{thr:split}
    Let $s_0, \ldots,s_{n-1}\in\mathbb{N}$  and $t_0,\ldots,t_{n-1}\in\mathbb{N}$ be sequences of natural numbers where each number is upper bounded by $s$ and $t$, respectively. Let $S=\sum_{j\in\interval{n}}{s_j}$ and $T=\sum_{j\in\interval{n}}{t_j}$. 
Then for any $k\in \mathbb{N}$, there is a partition of $\interval{n}$ into $k$ sets $B_0,\ldots,B_{k-1}$, such that for each $i$, the set $B_i$ consists of consecutive elements, and 

\begin{align*}
\sum_{j\in B_i}{s_j}\leq 2\left(\frac{S}{k}+s\right)\text{~~~and~~~ }&  \sum_{j\in B_i}{t_j}\leq 2\left(\frac{T}{k}+t\right).
\end{align*}
\end{claim}

Invoking \cref{thr:split} with $s=n\geq s_{i,j}$, $t=n\geq t_{i,j}$, $S=S_i$, and $T=T_i$, implies that for each $i\in [n]$ and $i'\in [k_i]$, the nodes inside each bucket $B_{i,i'}$ want to send/receive at most $4n$ messages, i.e., 
\begin{align*}
    \sum_{j\in {B_{i,i'}}}s_{i,j}& \leq2\left(\frac{S}{k}+s\right)\leq  2\left(\frac{S_i}{(S_i/s)} + s\right)=4s=4n\text{, and} \\
     \sum_{j\in {B_{i,i'}}}t_{i,j}& \leq 2\left(\frac{T}{k}+t\right) \leq 2\left(\frac{T_i}{(T_i/t)} + t\right)= 4t= 4n.
\end{align*}

\textit{Computing the assignment $\varphi$:} We first define the assignment $\varphi$ and then show how it becomes globally known.
Recall that the buckets of machine $p_i$ are numbered from $0$ to $k_i-1$ and define the following value for $i\in [n]$ and $i'\in\interval{k_i}$:
\begin{align*}
f(i,i')=\floor*{\left(i' + \sum_{i''<i}{k_{i''}}\right)/5}.
\end{align*}
Then, we define the assignment $\varphi$ to assign all nodes in bucket $B_{i,i'}$  to machine $p_{f(i,i')}$. 
Notice that this is a valid assignment because with $\sum_i S_i\leq2 n^2$ and $\sum_i T_i\leq 2n^2$ (due to $m^r\leq 2 n^2$) we obtain
\begin{align*}
f(i,i') 
< \sum_{0\leq i< n}{\frac{k_i}{5}}
= \frac{1}{5}\sum_{i}{\ceil{\max{\set{\frac{S_i}{n},\frac{T_i}{n}}}}}
\leq \frac{1}{5}\sum_{i}{\left(\frac{S_i}{n} + \frac{T_i}{n}+1\right)}
\leq\frac{1}{5}\cdot 5n=
n.
\end{align*}
Here, the first inequality follows from $i' < k_{i'}$.
Also, notice that each machine receives at most $5$ different buckets because at most five pairs $(i,i')$ are mapped to the same index by $f$.

Now, we want to make the assignment $\varphi$ globally known to all machines. To this end,  each machine $p_i$ broadcasts the number of its buckets, $k_i$. Thus, machine $p_i$ can compute $f(i,i')$ for each of its buckets $B_{i,i'}$. Then, for all $i'\in\interval{k_i}$, machine $p_i$ informs machine $p_{f(i,i')}$ about the smallest and the largest job number of a node in bucket $B_{i,i'}$. As the buckets $B_{i,1},\ldots,B_{i,k_i}$ are ordered (increasingly) by the jobs' indices for all $i\in [n]$, this information is sufficient for each machine to deduce which nodes are assigned to it in $\varphi$. In the last step, each machine broadcasts the messages that it has received, i.e., machine $p_{f(i,i')}$ broadcasts the smallest and largest job index of bucket $B_{i,i'}$ together with the index $i$,  and each machine can deduce the full assignment $\varphi$.

\textit{Executing round $r+1$:}
We now use \cref{thr:reassigningNodeRolesInJobs} to distribute the states $S_{i, j, r}$ from the trivial initial assignment $\varphi_s(i, j)=i$ to the globally known final assignment $\varphi_f=\varphi$.  Then, each machine executes the \ref{phase:writing} of round $r$ for each of its newly assigned nodes, where a message from $v_{i, j}$ to $v_{i', j}$ is sent from $p_{\varphi(i, j)}$ to $p_{\varphi(i', j)}$. This is possible since $\varphi$ is globally known.
Then, each machine executes the remainder of the protocol of its newly assigned nodes until after the \ref{phase:computation} of round $r+1$. Finally, the obtained states $S_{i,j,r+1}$ for round $r+1$ are re-distributed according to the trivial assignment by using  \cref{thr:reassigningNodeRolesInJobs} once more, with $\varphi_s=\varphi$ and $\varphi_f(i, j)=i$. 

\textbf{Correctness.}
For each $i\in\interval{n}$ and $j\in\interval{t}$ the machine $p_{f(i, j)}$ receives the state $S_{i,j,r}$ and executes the \ref{phase:writing} of round $r$,  the \ref{phase:reading} of round $r+1$, and the \ref{phase:computation} of round $r+1$ for node $v_{i,j}$. Thus, afterwards it holds the state $S_{i,j,r+1}$. Since this state is then sent back to $p_i$, the correctness follows.

\textbf{Round Complexity.}
The partitioning of each machine's nodes into buckets is done locally without communication. Broadcasting the number of buckets (the value of $k_i$) can be done in a single round.  
We next reason about the time complexity that is required to make the assignment $\varphi$ globally known.
The computation of $f(i, i')$ is done locally. Informing machine $p_{f(i, i')}$ about the smallest and largest job in the bucket $B_{i, i'}$ requires for each machine $p_i$ to send at most $t$ messages and to receive at most $5$ messages. Thus, by $\ceil{t / n}$ invocations of  \cref{thr:optimalRouting}, this step completes in $\bigO{\ceil{t / n}}$ rounds. Since each machine $p_i$ is assigned at most $5$ buckets, and for each bucket $B_{i', j}$ it broadcasts a constant number of elements (smallest and largest job index in it together with the identifier $i'$), this step completes in $\bigO{1}$ rounds.

The runtime is hence dominated by distributing the states via \cref{thr:reassigningNodeRolesInJobs}, which takes $\bigO{\ceil{M\cdot t/n}}$ rounds. All nodes in a bucket  send/receive at most $4n$ messages in total and each machine executes the sending/receiving phase for at most $5$ buckets, and thus these steps are done in $O(1)$ rounds by \cref{thr:optimalRouting}.
\end{proofof}

The next lemma deals with the general case, where total number of messages $m^r$ might be larger than $2n^2$.
\begin{lemma}[Scheduling of a round $r$.]\label{thr:mMessagesFromSqrtNJobs}
Given are $t$ jobs that are $M$-memory efficient, and given is a round number $r$. Assume that for each $i \in [n]$, $p_i$ holds $S_{i,j,r}$ for all $j\in [t]$. Then there exists a deterministic \CC algorithm which completes in $\bigO{\ceil{m^r / n^2 + M\cdot t / n}}$ rounds, at the end of which, for each $i \in [n]$, $p_i$ holds $S_{i,j,r+1}$ for all $j\in [t]$.
\end{lemma}
The proof of
\cref{thr:mMessagesFromSqrtNJobs}
uses the next claim to split all jobs into chunks that send smaller numbers of messages in order to apply \cref{thr:n2messagesFromSqrtNJobsIn1Round}.

\begin{claim}\label{thr:naryserach}
    Let $\mathcal{S}$ 
    be a non-empty (globally known) set of consecutive indices of size at most $n^c$ for some constant $c>0$ and let $x>0$. Each machine $p_i$ has a sequence of numbers $(s_{i, j})_{j\in \mathcal{S}}$ that are all upper bounded by $n$. There is a deterministic algorithm in the \CC model, which in $O(1)$ rounds finds the minimum index $j_0\in \mathcal{S}$ (if it exists) that satisfies 
		\begin{align} \label{eq:findingIndex}
		x\leq \sum_{\substack{j\in \mathcal{S},\\ j\leq j_0}}{\sum_{i=0}^{n - 1}{s_{i, j}}} \leq x + n^2.
		\end{align}
\end{claim}
We solve this problem in $c$ recurrent levels. On recursion level ${c'}$, which goes from $c$ down to $1$, we start with the search space $\mathcal{S}^{c'}$ of size $n^{c'}$ and finish with the search space $\mathcal{S}^{{c'} - 1}$ of size $n^{{c'} - 1}$. After each iteration, we maintain the invariant that if there exists the required $j_0$ then $j_0 - 1\in \mathcal{S}^{{c'} - 1}$ and that $\mathcal{S}^{{c'} - 1}$ is contiguous. We always maintain a search space of consecutive indices. 

Next, we explain the ${c'}$-th recursion level, and for that purpose assume that the  current search space $\mathcal{S}^{c'}$ is of size $n^{c'}$ for $0\leq {c'}\leq c$. If this is not the case, we append dummy indices to make $\mathcal{S}^{c'}$ of the size exactly $n^{c'}$.
To narrow down the search space, we compute $n$ prefix sums $S_{\ell_1}, \ldots, S_{\ell_{n}}$ where $S_{\ell_{i'}}$ sums up all values with index $j<\ell_{i'}$ of all machines. The indices $\ell_0=\min{\mathcal{S}},\ldots,\ell_{n}=\ell_0 + n^{c'}$ are equidistantly placed in $\mathcal{S}^{c'}$. Let $i'$ be the largest index such that $S_{\ell_{i'}} < x$. The new search space is formed by the indices $\mathcal{S}^{{c'} - 1}=\rightopeninterval{\ell_{i'}, \ell_{i' + 1}}$.

After the last recursion level we obtain singleton search space $\mathcal{S}^0$. We return $j_0$ as that value plus one if it is less than $n^c$. Otherwise, respond that the required $j_0$ does not exist.

\begin{proofof}[thr:naryserach]    \textbf{Algorithm:}
    Initially we may assume that the search space $\mathcal{S}$ is of size exactly $n^c$.  If this is not the case, we append dummy indices to the end of $\mathcal{S}$, in other words we add the indices $\set{\max{\mathcal{S}} + 1, \max{\mathcal{S}} + 2, \dots, \max{\mathcal{S}} + n^c - \size{\mathcal{S}}}$ to $\mathcal{S}$, obtaining the range of indices $[\min{\mathcal{S}},\dots,\min{\mathcal{S}+n^c-1}]$.
    We proceed in $c$ recursion levels, in each of which we decrease the size of the search space by a factor of $n$, while always maintaining a search space of consecutive indices.
    
    Consider iteration ${c'}$ with the search space $\mathcal{S}^{c'}$. Let $\ell_0$ be the smallest index in $\mathcal{S}^{c'}$ and for $1\leq i'\leq n$ let $\ell_{i'}=\ell_0+ i'\cdot n^{{c'}-1}$.
    Now, each machine $p_i$ builds prefix sums $S^i_{\ell_1}, \ldots,S^i_{\ell_{n}}$ of its own numbers, that is \[S^i_{\ell_{i'}}=\sum_{\substack{j\in  0\leq j < \ell_{i'}}} s_{i,j}.\] Then, all machines send their computed prefix sum corresponding to $\ell_{i' + 1}$  to machine $p_{i'}$ which sums up all received prefixes, that is, afterwards machine $p_{i'}$ holds $S_{\ell_{i' + 1}}=\sum_{i\in\interval{n}} S^i_{\ell_{i' + 1}}$. In a second round of communication $S_{\ell_1}, \ldots, S_{\ell_n}$  are broadcasted and every node can determine the new search space $\mathcal{S}^{{c'} - 1}=\rightopeninterval{\ell_{i'}, \ell_{i'+1}}$ where $i'$ is the largest number such that the prefix sums $S_{\ell_{i'}}$ add up to less than $x$.
    After $c$ levels of recursion the search space consists of a single index $\ell$. We return $j_0=\ell + 1$.
    \textbf{Correctness:}
    By induction, before level ${c'}$ the search space size is $\size{\mathcal{S}^{c'}}=n^{c'}$ and the largest index $\ell$ such that $S_{\ell} < x$ belongs to $\mathcal{S}^{c'}$. Thus, after $c$ levels, the search space is a singleton $\ell$. This means that $x\leq S_{j_0}$ in case $j_0 < n^c$. In case we return that $j_0$ does not exist, it holds that $\ell=n^c - 1$, and so the sum of all $s_{i, j}$ is indeed below $x$.
    
    As each $s_{i,j}$ is upper bounded by $n$, we obtain $\sum_{i\in [n]}s_{i,j_0}\leq n^2$, and as $S_{j_0-1}=\sum_{j<j_0}\sum_{i\in [n]}s_{i,j}\leq x$,  we obtain the claimed upper bound in Eq(\ref{eq:findingIndex}). 
    
    \textbf{Round complexity:}
    As all $s_{i,j}$ are upper bounded by $n$ and $t$ is polynomial in $n$, all numbers can be send in $O(\log n)$-bit messages. Each recursion level can be implemented in $O(1)$ rounds, thus we need $O(c)=O(1)$ rounds in total.
\end{proofof}

We continue with the proof of \cref{thr:mMessagesFromSqrtNJobs}.

\begin{proofof}[thr:mMessagesFromSqrtNJobs]
\textbf{Algorithm.} A short pseudocode is given in \Cref{alg:moreN2}. 
\begin{algorithm}
\caption{Scheduling of a round.}
\label{alg:moreN2}
\begin{algorithmic}[1]
    \State{Split jobs into chunks $J_1, J_2, \dots, J_k$.}
    \For{each chunk $J_{k'}$}
        \State{Apply \cref{alg:atMostN2} on $J_{k'}$.}
    \EndFor
\end{algorithmic}
\end{algorithm}

We use \cref{thr:naryserach} to split the jobs into $k=O(\ceil{m^r/n^2})$ chunks $J_1,\ldots,J_k$, such that the jobs in each chunk send at most $2n^2$ messages in round $r$ over all of their nodes.  Then, we iteratively apply \cref{thr:n2messagesFromSqrtNJobsIn1Round} on each chunk to progress each job to the next round. 

\textit{Forming chunks}: 
First, each machine $p_i$, for each job $j$, uses $S_{i, j, r}$ to locally compute the number of messages $s_{i, j}$ that node $v_{i, j}$ sends in round $r$. 
Assume that chunks $J_1,\ldots,J_{k'-1}$ 
have been formed and let $\mathcal{S}=[t]\setminus (J_1\cup\cdots\cup J_{k'-1})$. 
We apply \cref{thr:naryserach} 
with the index set $\mathcal{S}$, where machine $p_i$ holds the sequence $(s_{i,j})_{j\in \mathcal{S}}$, and with $x=n^2$. If we find $j_0$, by the guarantee of \cref{thr:naryserach}, we obtain that all jobs in a chunk $J_{k'}$, for ${k'}\neq k$, send at least $n^2$ messages and at most $2\cdot n^2$ messages in round $r$. The jobs in chunk $k$  send at most $2n^2$ messages. Otherwise, if we do not find $j_0$, the nodes of the jobs in $\mathcal{S}$ send less than $2n^2$ messages in round $r$, so we obtain the last chunk and set $J_k=J_{k'}=\mathcal{S}$. We thus have $k\leq \ceil{m^r/n^2}$.

\textit{Executing round $r + 1$:} Since, by construction, the jobs in each chunk send at most $2n^2$ messages, we can iteratively apply \cref{thr:n2messagesFromSqrtNJobsIn1Round} on the chunks.

\textbf{Round complexity.}
We split the jobs into at most $k=O(\ceil{m^r/n})$ chunks, where forming each chunk can be done in $O(1)$ rounds by \cref{thr:naryserach}. The invocation of \cref{thr:n2messagesFromSqrtNJobsIn1Round} on chunk $J_{k'}$  takes $\bigO{\ceil{M  \size{J_{k'}} / n}}$ rounds per chunk. Thus, the round complexity of the algorithm is \begin{align*}
    \bigO{1} + \sum_{k' = 1}^{k}\bigO{\ceil{M \cdot \size{J_{k'}} / n}}
    & =\bigO*{k + M\cdot\sum_{k' = 1}^{k}{\size{J_{k'}}} / n}
     =\bigO*{\ceil{m^r / n^2 + M\cdot t / n}}. & & \qedhere
\end{align*}
\end{proofof}

Finally, we use \Cref{thr:n2messagesFromSqrtNJobsIn1Round,thr:mMessagesFromSqrtNJobs} to obtain the near-optimal scheduling of \Cref{thr:optimalShedulingOfSmallJobs}.

\deterministic*
\begin{proofof}[thr:optimalShedulingOfSmallJobs]
We repeatedly apply \cref{thr:mMessagesFromSqrtNJobs} until all jobs terminate.
First, each machine $p_i$ reads the input for each node $v_{i,j}$ for each $j\in [t]$, and executes  the \ref{phase:computation} of round $r=0$, as a result of which it holds the state $S_{i, j, 0}$ for each of its nodes. Then, we split the execution into epochs, where in epoch $r$ all jobs move from the \ref{phase:computation} of round $r$ to the \ref{phase:computation} of round $r+1$. A single epoch is implemented via \cref{thr:mMessagesFromSqrtNJobs} in $\bigO{\ceil{m^r / n^2 + M\cdot t / n}}$ rounds. 
After the epoch $r=\dilation-1$,  all machines compute the outputs given the respective terminating state $S_{i, j, \dilation - 1}$ of each of its nodes. 

\textbf{Round complexity.}
The pre-processing in round $r=0$ and the  post-processing in the last round $r=\dilation - 1$ is done locally and does not require communication. Due to \cref{thr:mMessagesFromSqrtNJobs}, the round complexity of executing round $r$ for all jobs is $\bigO{\ceil{m^r / n^2 + M\cdot t / n}}$, where $m^r$ is the number of messages sent in round $r$. Since $\sum_{r=0}^{\dilation-1}m^r/n^2=\bandwidth$, we obtain the overall  round complexity by 
\begin{align*}
    \sum_{r\in\interval{\dilation}}{\bigO{\ceil{m^r / n^2 + M\cdot t / n}}}
    & = \bigO*{\bandwidth + \dilation \cdot \ceil{M\cdot t / n}}. & & \qedhere
\end{align*}
\end{proofof}

\section{Randomized Scheduling}\label{sec:randomized}
In this section we show and compare the two approaches for randomized scheduling: random shuffling (\cref{subsec:randomPermutation}) and random delaying (\cref{subsec:delay}). In contrast to \cref{thr:optimalShedulingOfSmallJobs}, the results in this section do not require the jobs to be memory efficient.

\subsection{\label{subsec:randomPermutation}Scheduling through Random Shuffling}

In this subsection we use random shuffling to schedule I/O efficient jobs and we obtain the following theorem.
{\renewcommand\footnote[1]{}\shuffle}

The definition of an I/O efficient job as used in \cref{thr:shuffle} is as follows.
\begin{definition}[I/O efficient job]
An \emph{I/O efficient job} is a job 
where each node receives and produces at most $\bigO{n\log{n}}$ bits of input and output. 
\end{definition}

\paragraph{Algorithm.}
The high level overview of the algorithm for \cref{thr:shuffle} (see \cref{alg:shuffle}) consists of three steps: \texttt{Input Shuffling}, \texttt{Execution}, and \texttt{Output Unshuffling}.
\begin{algorithm}\caption{Scheduling of I/O efficient job.}\label{alg:shuffle}
\begin{algorithmic}[1]
    \State{\label{step:shuffle:input}\texttt{Input Shuffling}}
    \State{\label{step:shuffle:execute}\texttt{Execution}:} Run $\dilation$ many phases, where in phase $r$ each machine $p_i$ runs the protocol for its nodes $\{ v_{\pi_j^{-1}(i),j} \mid j\in [t] \}$, and messages are routed  via \cref{thr:optimalRouting}. 
    \State{\label{step:shuffle:output}\texttt{Output Unshuffling}}
\end{algorithmic}
\end{algorithm}

\texttt{Input Shuffling:}
We iterate sequentially through the jobs. For each job, a leader machine, say, $p_0$, generates a random uniform permutation $\pi_j\colon\interval{n}\mapsto\interval{n}$. The permutation becomes globally known within two rounds by having $p_0$ send $\pi_j(i)$ to each $p_i$ and then each $p_i$ broadcasts $\pi_j(i)$ to all machines. In the last round of this subroutine, each machine $p_i$ sends the input of $v_{i,j}$ to machine $p_{\pi_j(i)}$. A single round is sufficient because the job is I/O efficient. Thus, at the end, machine $p_i$ holds the state of the nodes $v_{\pi_j^{-1}(i), j}$ for all $j \in [t]$. We call this subroutine \texttt{Input Shuffling}. 

\texttt{Execution:}
In \dilation many phases we progress each job by one round. That is, each machine $p_i$ performs all actions of the nodes that it holds, which are $v_{\pi^{-1}_j(i),j}$ for all $j \in [t]$. In order to  use \cref{thr:optimalRouting} efficiently for each phase $r$, the machines need to compute a bound on the number of messages that any of them sends or receives in phase $r$. To this end, the machines jointly compute 
the value of $m^r=\sum_{j\in [t]}\sum_{i \in [n]}{s^r_{i,j}}$, where $s^r_{i,j}$ is the number of messages that node $v_{i,j}$ sends in round $r$. They do this by having each machine $p_i$ send $\sum_{j \in [t]}{s^r_{\pi^{-1}_j(i),j}}$ to a leader machine, say, $p_0$, which then sums these values and broadcasts their sum $m^r$. 
That is, $m^r$ is the total number of messages sent by all nodes in all jobs in round $r$, and we show that for each $i\in[n]$, $O(m^r/n+n\log n)$ is
a bound on $\sum_{j \in [t]}{s^r_{\pi^{-1}_j(i),j}}$ ($\sum_{j \in [t]}{t^r_{\pi^{-1}_j(i),j}}$), which is the number of messages that machine $p_i$ has to send (receive) in phase $r$, to be used when invoking \cref{thr:optimalRouting}. 

\texttt{Output Unshuffling:} At the end, after each machine executes the protocols until they finish, we use a single round of communication for each job to unshuffle the outputs according to $\pi_j^{-1}$. At the end of this \texttt{Output Unshuffling} subroutine, machine $p_i$ holds the 
output
$v_{i, j}$ for all $j \in [t]$. 
This finishes the description of the algorithm.
\medskip

In the following lemma, we bound the number of messages that each machine has to send/receive in one phase by $X=O(m^r/n+n\cdot\log n)$.
\begin{restatable}{lemma}{boundPerPhase}\label{lem:boundPerPhase}
Consider $t$ jobs and a set of permutations $\{\pi_j\}_{j\in\interval{t}}$ generated uniformly at random and let ${S}=\max_{i\in\interval{n}}\sum_{j\in\interval{t}}{s^r_{\pi^{-1}_j(i), j}}$ and ${R}= \max_{i\in\interval{n}}\sum_{j\in\interval{t}}{t^r_{\pi^{-1}_j(i), j}}$.
Then, \whp, it holds that $X=\max\{{S}, {R}\}=O(m^r/n+n\log n)$, where $m^r=\sum_{i\in\interval{n}}
\sum_{j\in\interval{t}}{s^r_{i, j}}$. 
\end{restatable}
\begin{proofof}[lem:boundPerPhase]
    Let $c\geq 1$ be arbitrary large constant.     Denote by $S^r_{i, j}=\sum_{i'\in [n]}s^r_{i',j}\cdot \Indicator_{\pi_j(i')=i}$ the random variable whose value is the number of messages sent by machine $p_i$ for job $j$ (note that there is a single $i'=\pi^{-1}_j(i)$ for which $i=\pi_j(i')$, but this $i'$ is also a random variable). These variables are bounded by $n$ and are independent for different $j$. Denote by $S^r_i=\sum_{j\in\interval{t}}{S^r_{i, j}} / n$ the random variable whose value is the total number of messages machine $p_i$ sends normalized by $n$. Denote $c' = c + 2$. We  show that the normalized number of messages machine $p_i$ sends is bounded as $S^r_i\leq 3m^r/n^2 + 2 c' \ln{n}$, with probability at least $1 - n^{c'}$.
    
    First, we note that the expected normalized number of messages machine $p_i$ sends is:
    \begin{align*}
                \E*{\sum_{j\in\interval{t}}S^r_{i, j} / n}
        & =\sum_{j\in\interval{t}}\sum_{i'\in [n]}s^r_{i', j}\cdot \E*{\Indicator_{\pi_j(i')=i} / n} \\
        & ={\sum_{j\in\interval{t}}\sum_{i'\in [n]}s^r_{i', j} / n^2}
        = m^r / n^2,
    \end{align*}
where the first equality holds due to the linearity of expectation, the second one holds since $\pi_j$ is sampled uniformly and the last one is due to the definition of $m^r$.

    Since for different $j$, the variables ${S_{i, j}^r}/{n}$ are independent, we use \cref{thr:relativeHoeffding} (Hoeffding Bound) with a relative error $\epsilon > 0$, which we later optimize, to bound the probability that a machine has too many messages to send.    \begin{align*}
        \Pr\left[S^r_{i} > \left(1+\epsilon\right) \frac{m^r}{n^2}\right] 
         = \Pr\left[
            \sum_{j\in\interval{t}} \frac{S^r_{i, j}}{n}
            > 
            \left(1+\epsilon\right) \frac{m^r}{n^2}
        \right] 
         <
        e^{-\frac{\epsilon^2 m^r}{(2+\epsilon)n^2}}.
    \end{align*}
    
            If $m^r\geq c'\cdot n^2 \ln n$, then for $\epsilon=2$ we have that $e^{-{\epsilon^2 m^r}/{((2+\epsilon)n^2)}}\leq e^{-c'\ln n}=n^{-c'}$. In other words, \whp $3{m^r}/{n^2}=\bigO{{m^r}/{n^2}}$ rounds are sufficient for machine $p_i$ for sending all required messages on round $r$.
        Otherwise, we have $m^r < c'\cdot n^2\ln n$. In this case, for $\epsilon=2c'\cdot {n^2\ln(n)}/{m^r}\geq 2$ we get that $e^{-{\epsilon^2 m^r}/{((2+\epsilon)n^2)}}= e^{-{2\cdot c'\ln{(n)}}/{({2}/{\epsilon}+1)}}\leq n^{-c'}$. In other words, \whp  $(1+2c'\ln{(n)}\cdot {n^2}/{m^r}){m^r}/{n^2}={m^r}/{n^2}+2c'\cdot \ln n=\bigO{{m^r}/{n^2}+\log{n}}$ rounds are sufficient for machine $p_i$ for sending all of its required on round $r$. We conclude that $\Pr\left[
                        S^r_i
            > 
            3{m^r}/{n^2}+2c'\ln n
        \right] < n^{-c'}$.
    
    Denote by $T^r_i$ the random variable whose value is the number of messages received by  machine $p_i$ normalized by $n$. By the same approach, we show that \[\Pr\left[
        T^r_i
        > 
        3\frac{m^r}{n^2}+2c'\ln n
    \right] < n^{-c'}.\]
    
    By a union bound over $S^r_i$, $T^r_j$ for all $i \in [n]$, we obtain that for some $i$ one of the event $S^r_i >  3({m^r}/{n^2}+2(c + 2)\ln{n})$, $T^r_i >  3({m^r}/{n^2}+2(c + 2)\ln{n})$ happens with probability at most $2n^{-c' + 1}\leq n^{-c}$ for $n\geq 2$. Notice, that ${S}=\max_{i\in\interval{n}}{S_i\cdot n}$ and ${R}=\max_{i\in\interval{n}}{T_i\cdot n}$, thus $X=\max\set{{S}, {R}}=O(m^r/n+n\log n)$ \whp 
\end{proofof}

With an upper bound at hand, on the number of messages that each machine sends or receives in phase $r$, we can prove that \cref{alg:shuffle} satisfies the statement of \cref{thr:shuffle}.
\begin{proofof}[thr:shuffle]
We prove the correctness and bound the runtime of the presented algorithm (see \cref{alg:shuffle}). 

\textbf{Correctness:} 
After the \texttt{Input Shuffling} subroutine (\cref{step:shuffle:input}), the input for node $v_{i, j}$ is stored on machine $p_{\pi_j(i)}$. For each phase $r \in [\dilation]$, we invoke \cref{thr:optimalRouting} with the computed value $X$, which is \whp a bound the number of messages that each machine sends or receives. Thus, \whp this invocation succeeds. Since $\dilation=\bigO{n}$, a union bound over all phases gives that at the end of the \texttt{Execution} subroutine, each machine $p_i$ holds the outputs of all nodes $v_{\pi^{-1}(i),j}$ for each $j \in [t]$. After \texttt{Output Unshuffling}, machine $p_i$ holds the output for node $v_{i, j}$ for each job $j\in [t]$. 

\textbf{Round Complexity:}
The initial \texttt{Input Shuffling} (\cref{step:shuffle:input}) and the \texttt{Output Unshuffling} at the end of the algorithm (\cref{step:shuffle:output}) complete with $t$ rounds each. For each phase $r$ in the  \texttt{Execution} part of the algorithm, computing $m^r$ is done in 2 rounds. By \cref{lem:boundPerPhase}, $X=\bigO{m^r/n+n\log n}$ is a bound on $\sum_{j\in [t]}{s^r_{\pi^{-1}(i),j}}$ and $\sum_{j\in [t]}{t^r_{\pi^{-1}(i),j}}$, which are the number of messages that  machine $p_i$ sends and receives in phase $r$, respectively, for all $i \in [n]$. Thus, invoking \cref{thr:optimalRouting}
completes in $\bigO{m^r / n^2 + \log{n}}$ rounds, \whp. Thus, the overall round complexity of the algorithm is 
\begin{align*}
\bigO{t + \sum_{r\in\interval{\dilation}}{(m^r / n^2 + \log{n})}}
= \bigO{t + \sum_{j\in\interval{t}}{m_j} / n^2 + \dilation\cdot\log{n}}
\\
= \bigO{t + \bandwidth + \dilation\cdot\log{n}}
. & 
\qedhere
\end{align*}
\end{proofof}

\subsection{\label{subsec:delay}Scheduling through Random Delays}

In this subsection we show how to use random delays approach introduced in \cite{LeightonMR94} to schedule round efficient jobs. \begin{theorem}\label{thr:delay}
There is a randomized algorithm in the \CC model, which schedules $t$ jobs
\[\bigO{\capacity  +  \dilation\cdot\log{n}+{t/n}}\] rounds, \whp, given an upper bound on the value of \capacity, and in   \[O\big(\capacity +\log\capacity\cdot ( \dilation\cdot\log{n}+t/n)\big)\] rounds, \whp, if such a bound is not known.
\end{theorem}
In the algorithm, job $j\in [t]$ is executed with a delay $D_j$ that is chosen uniformly at random from $[D]$, where $D=\floor{\capacity / \ln{n}}$.
In the crucial step of the proof, we use a Hoeffding Bound to show that this random delay implies that each node has to send and receive at most $X=O(\capacity\cdot n / D)$ messages per round in all jobs combined. The claim then follows by routing all messages of a single round with  Lenzen's routing scheme (\cref{thr:optimalRouting}). 
This approach uses that all nodes know a bound on \capacity, which can be removed at the cost of a logarithmic factor with a standard \emph{doubling}-technique. 

\textbf{Algorithm:} We describe the algorithm for the case where $\capacity$ is known. The algorithm consists of initializing part \texttt{Sample Delays}, followed by the actual \texttt{Execution} part. Let $D=\floor{\capacity / \ln{n}}$.

\texttt{Sample Delays:}
 We start by generating a random delay $D_j$ for each job $j$ and broadcasting it. For this, a leader node, say, $p_0$, samples  a delay $D_j$ uniformly at random from $\interval{D}$ independently for each job $j$. Notice, that in the special case $D\leq 1$ (which happens when $\capacity< 2\ln n$), the delays are actually degenerated to the deterministic $D_j=0$.
\texttt{Execution ($O(D+\dilation)$ phases):}
In \emph{phase $r$} we progress each job $j$ (for which $r \geq D_j$ holds) from round $r - D_j$ to round $r - D_j + 1$. Each machine $p_{i}$ executes the protocol of round $r - D_j$ for job $j$. To deliver the messages efficiently, we use the algorithm from \cref{thr:optimalRouting}, which requires the bound $X$ on $\max_{i\in\interval{n}}\set{\sum_{j\in\interval{n}}s_{i, j}^{r - D_j}, \sum_{j\in\interval{n}}t_{i, j}^{r - D_j}}$, the number of messages machine sends or receives. If $\capacity < 2\ln{n}$, the number of messages to send or receive is clearly bounded by $\bigO{n\log{n}}$. In the general case, we show that this bound is $\bigO{\capacity\cdot n / D}$ \whp.

\texttt{Doubling:}
To remove the requirement on the knowledge of \capacity, we use a standard doubling technique. We try to run the algorithm until success while doubling the estimation of \capacity in each attempt, starting from a guess of $\capacity=1$. The algorithm detects failure when the algorithm from \cref{thr:optimalRouting} fails.

\begin{algorithm}\caption{Scheduling of jobs.}\label{alg:delay}
\begin{algorithmic}[1]
    \State{\label{step:delay:sample}\texttt{Sample delays}:
    Independently UAR pick $D_j\in[D]$ and broadcast the values}
    \State{\texttt{Execution }: 
    Run $O(D + \dilation)$ phases, where  in phase $r$ progress each job $j$ that satisfies $r \geq D_j$ by one round where the messages of all jobs are routed with \cref{thr:optimalRouting}.
    }
\end{algorithmic}
\end{algorithm}

In the proof of the following lemma, we bound the number of messages that each machine has to send/receive in one phase by $X=\bigO{\capacity\cdot n / D}$.

\begin{lemma}\label{thr:delay:messages}
    Given $t$ jobs and a set of delays $\{D_j\}_{j\in\interval{t}}$ sampled uniformly at random from $\interval{D}$ for $D=\floor{\capacity / \ln{n}} \geq 1$, let ${S}=\max_{i\in\interval{n}}{\sum_{j\in\interval{t}\colon r\geq D_j}{s^{r - D_j}_{i, j}}}$, ${R}=\max_{i\in\interval{n}}{\sum_{j\in\interval{t} \colon r\geq D_j}{t^{r - D_j}_{i, j}}}$, and  $X=\max\set{{S}, {R}}$.

    Then, \whp, it holds that $X=O(\capacity\cdot n / D)$, where $m^r=\sum_{i\in\interval{n}} \sum_{j\in\interval{t}}{s^r_{i, j}}$. 
\end{lemma}
\begin{proofof}
    Let $c\geq 1$ be arbitrary large constant. Denote by $S_{i, j}=\sum_{r'\in\interval{\dilation}}s_{i, j}^{r'}\cdot \Indicator_{D_j+r'=r}$ the random variable whose value is the number of messages sent by machine $p_i$ for job $j$ on round $r$. These variables are independent for different values of $j$, as the delays $D_j$ are independent. They are also bounded by $n$, which means that the variables $S_{i, j}/n$ are also independent and belong to $\interval{0, 1}$. Denote by $S_i=\sum_{j\in\interval{t}}{S_{i,j}} / n$ the random variable whose value is the number of messages sent by machine $p_i$, normalized by $n$. Denote $c'=c+2$.  We show that the normalized number of messages machine $p_i$ sends is bounded as $S_i^r \leq (1 + 2\cdot c')\capacity / D$ with probability at least $1 - n^{c'}$.
    
    First, we note that the expected normalized number of messages machine $p_i$ sends is:
    
    \begin{align*}
        \E{S_i}
        = \E*{{\sum_{j\in\interval{t}}{S_{i, j}}}/{n}} 
        & = \frac{\sum_{j\in\interval{t}}\sum_{r'\in\interval{\dilation}}s_{i, j}^{r'}\cdot \E{\Indicator_{D_j+r'=r}}}{n} & \\
        & \leq \frac{\sum_{j\in\interval{n}}\sum_{r'\in\interval{\dilation}}s_{i, j}^{r'}}{D\cdot n}
        =\frac{\capacity}{D},
    \end{align*}
    where the second transition is due to the linearity of expectation, the third follows from delays being uniformly selected and the last one is due to the definition of $\capacity$.
    
    Since $D_j$ are independent, we use \cref{thr:relativeHoeffding} (Hoeffding Bound) with $\epsilon=2\cdot c'$, we bound the probability of $S_i$ being larger than the expected value by
    \begin{align*}
        \Pr[S_i \geq (1+2\cdot c')(\capacity/D) ]
        & =\Pr[\sum_{j\in\interval{t}}{(S_{i, j}/n)}\geq (1+2\cdot c')(\capacity/D)]\\
         & \leq e^{-\frac{(2\cdot c')^2}{2+2\cdot c'}\frac{\capacity}{D}}
        =e^{-\frac{4c'^2}{2+2c'}\ln n}
       \leq  n^{-c'}       ,
    \end{align*}
    where the second transition is due to \cref{thr:relativeHoeffding}  and the third is due to the selection of $D\leq\frac{\capacity}{\ln{n}}$.
    
    Denote by $T_i$ the random variable whose value is the number of messages received by machine $p_i$ on round $r$ normalized by $n$. Using a similar approach, it holds that \[\Pr\left[T_i \geq (1 + 2\cdot c')\capacity / D \right]\leq n^{-c'}.\]
    By a union bound over all $S_i$ and $T_i$, we obtain that for some $i$, the probability that $S_i$ or $T_i$ are more than $\capacity / X$ is bounded by $n^{-c}$. Since ${S}=\max_{i\in\interval{n}}{S^r_{i}}$, ${R}=\max_{i\in\interval{n}}{T^r_{i}}$, and $X=\max\set{{S}, {R}}$ it \whp holds that $X=O(\capacity\cdot n / D)$.
\end{proofof}

The following simple routing primitives are used in the random-delay based algorithm of \cref{thr:delay}.
\begin{definition}{({Multiple broadcast problem.})}
Each machine $p_i\in V$ is given a set $M_i$ of $m_i$ messages of size \bigO{\log n} bits each. The goal is to deliver each message to all the machines. 
\end{definition}

\begin{lemma}\label{thr:multipleBroadcast}
There is an algorithm in the \CC model, which solves the multiple broadcast problem in $O\big(\ceil{{\sum_{i\in\interval{n}}m_i}/{n}}\big)$ rounds.
\end{lemma}
\begin{proofof}

The pseudocode is given in \cref{alg:optimalBroadcast}. First, on \cref{step:broadcast:count}, each machine $p_i$ broadcasts $m_i$, the number of messages it has. Given the information it receives, the machine $p_i$ locally computes $y_i=\sum_{i'=0}^{i - 1}{m_{i'}}$, the number of messages the machines with preceding identifiers $i'<i$ have. This allows each machine to compute indices of its messages in the global numbering. We split the execution into $\ceil{{\sum_{i\in\interval{n}}m_i}/{n}}=\ceil{y_n / n}$ phases. On phase $k$, a batch of messages with indices $[k\cdot n, \min\set{(k+1)\cdot n, \sum_{i\in\interval{n}}m_i})$ are broadcasted in two rounds. In the first round, the $i'$-th message of the current batch (e.g. the message number $k\cdot n + i'$) is sent to machine $p_{i'}$ (\cref{loop:broadcast:send}). In the second round, each machine broadcasts the message it received in the previous round (\cref{loop:broadcast:broadcast}). 

\begin{algorithm}
\caption{Multiple broadcasts.}
\label{alg:optimalBroadcast}
\begin{algorithmic}[1]
    \State{\label{step:broadcast:count}Each machine $p_i$ broadcasts $m_i$.}
    \State{\label{step:broadcast:offset}Each machine $p_i$ locally computes its $y_i=\sum_{i'=0}^{i - 1}{m_{i'}}$. 
    }
    \For{\label{loop:broadcast:phases}$k\gets 0$ to $\floor{{y_n}/{n}}$}
        \State{\label{loop:broadcast:send}For each $i'\in\interval{n}$ message number $k\cdot n + i'$ in global numbering is sent to the machine $p_{i'}$.}
        \State{\label{loop:broadcast:broadcast}Each $p_{i'}$ broadcasts the message it receives.}
    \EndFor
\end{algorithmic}
\end{algorithm}

In the first round of each phase, at most one message is received by each machine, in particular only $1$ message between any pair of machines. In the second round of each phase, each machine sends at most $1$ message to each other machine. Hence, the entire execution completes in $\bigO{\ceil{{\sum_{i=0}^{n - 1}m_i}/{n}}}$ rounds. 
\end{proofof}

\begin{proofof}[thr:delay]
We prove the correctness and bound the runtime for the aforementioned algorithm (\cref{alg:delay}).

First, in the special case $\capacity < 2\ln{n}$,
the number of messages each machine has to send over the entire execution for all jobs combined and in particular in each round is bounded by $2\cdot n\ln{n}$. Thus, a straightforward execution of one round of all jobs with \cref{thr:optimalRouting} completes in $2\ln{n}$ rounds, and the entire execution takes $\bigO{\dilation\cdot\log{n}}$ rounds. From now on we assume $D\geq 2$.

\textbf{Correctness.}
In each phase $r\in\interval{D + \dilation}$, we invoke \cref{thr:optimalRouting} with a bound of $X=\bigO{\capacity / D}$, which due to \cref{thr:delay:messages} bounds \whp the number of messages each node sends or receives. Thus, due to the union bound over $n$ rounds, all of them succeed \whp.

\textbf{Round complexity.}
Broadcasting $t$ values during \texttt{Sample Delay} (\cref{step:delay:sample}) takes $\bigO{\ceil{t / n}}$ rounds by \cref{thr:multipleBroadcast}. For each phase $r\in\interval{D + \dilation}$, by \cref{thr:delay:messages} $X=\bigO{\capacity\cdot n / D}$ is a bound on the number of messages that machine $p_i$ sends and receives in phase $r$ for each $i\in\interval{n}$ \whp and by applying union bound over the $D + \dilation=\bigO{\poly{(n)}}$ rounds, this holds on each round \whp. Thus, invoking the algorithm from  \cref{thr:optimalRouting} completes in $\bigO{\ceil{\capacity / D}}=\bigO{\capacity / D}$. Thus, overall, for $D=\floor{\capacity / \ln{n}}$ the algorithm terminates in $\bigO{\ceil{{t}/{n}}} + (\dilation + D)\cdot \bigO{{\capacity}/{D}} = \bigO{t / n + \dilation \log{n} + \capacity}$ rounds \whp

\textbf{Doubling.}
Since the algorithm succeeds \whp when our estimate is at least equal to the value of \capacity, we finish within $\bigO{\log{\capacity}}$ attempts. Thus, \whp, the round complexity of this approach is 
$
\sum_{\kappa = 0}^{\bigO{\log{\capacity}}}\bigO{t / n + 2^\kappa + \dilation\cdot\log{n}}
= \bigO{\capacity + \log{\capacity}\cdot (t / n + \dilation\cdot\log{n})}
$.
\end{proofof}

\section{Applications: MIS \& Pointer Jumping}
\label{sec:applications}

In this section we apply the scheduling algorithms developed in \cref{sec:deterministic,sec:randomized} on protocols which solve MIS (\cref{subsec:MIS}) and Pointer Jumping (\cref{subsec:pointerJumping}).
We analyze the round complexity of the developed algorithms.

\subsection{Maximal Independent Set}\label{subsec:MIS}

A \emph{maximal independent set (MIS)} of a graph $G=(V, E)$ is a subset of nodes $M\subseteq V$ such that no two nodes in $M$ are connected by an edge and adding any node to $M$ would break this property.
In this subsection, we show that we can efficiently solve multiple MIS instances using our scheduling algorithm from \cref{thr:shuffle}. 

\MISAmortized

To prove our result, we prove that the MIS protocol for the \CC model given in \cite{GhaffariGKMR18}, which completes in  $O(\log\log\Delta)$ rounds, uses $O(n^2)$ messages in all rounds combined, which we state as follows.
\begin{theorem}[Analysis of the MIS protocol of \protect{\cite[Theorem 1.1]{GhaffariGKMR18}}]
There is a randomized MIS protocol in the \CCL model which completes in   $\bigO{\log\log\Delta}$ rounds and sends $\bigO{n^2}$ messages, \whp
\label{thr:MIS}
\end{theorem}

Given \cref{thr:MIS}, we prove \cref{thr:MISAmortized} as follows.
\begin{proofof}[thr:MISAmortized]
By \cref{thr:MIS}, a set of $t$ jobs of the MIS protocol of \cite{GhaffariGKMR18} have $\dilation=O(\log\log\Delta)$ and $\bandwidth=\frac{t\cdot n^2}{n^2}=t$, \whp. By \cref{thr:shuffle}, \whp, we can schedule the $t$ jobs in a number of rounds bounded by $\bigO{t+\bandwidth+\dilation\cdot\log n}=\bigO{t+\log\log\Delta\log n}$.
\end{proofof}

\paragraph{Remark.}
\cref{thr:MISAmortized} also shows that the random-shuffling approach may be more efficient than random-delays.
In the MIS protocol of \cite{GhaffariGKMR18} which we use here, a leader node is used for collecting some of the edges of the graph. Potentially, since the leader node may receive $\bigO{n}$ messages during the $\bigO{\log\log\Delta}$ rounds of the protocol, applying the random-delay scheduling of \cref{thr:delay} on $t$ such MIS jobs results in a complexity of $\bigO{t\log\log\Delta + \log\log\Delta \log{n}}$ rounds. This run-time is asymptotically worse than the one obtained by the algorithm from \cref{thr:MISAmortized} for $t=\bigOmega{\log{n}}$. Moreover, it is no better then the na\"ive execution of the protocol multiple times one after another.

It remains to prove \cref{thr:MIS}.
\begin{proofof}[thr:MIS]
The correctness and the round complexity follow from \cite{GhaffariGKMR18}. We analyze the message complexity of the protocol.
For this we must describe the protocol, which returns a set $M\subseteq V$, initially empty. 

\paragraph{The Protocol (See \cref{alg:MISnew}).}~

\textbf{Random ranking:} First, a leader node $v^*$ generates a uniform random permutation $\pi\colon\interval{n}\mapsto\interval{n}$ and makes it globally known within 2 rounds by sending each node $v_i$ the value of $\pi(i)$ which $v_i$ then broadcasts to everyone. The value $\pi(i)$ is called the \emph{rank} of $v_i$ and does not change during the algorithm.

\textbf{Degree reduction by simulating greedy steps:} The second part of the protocol is a loop, which, as shown in \cite[Theorem 1.1]{GhaffariGKMR18},  uses $O(\log\log\Delta)$ iterations \whp,  to reduce the maximum degree of \emph{active} nodes to $\Delta'=\min\set{\Delta,\polylog{n}}$.

\begin{algorithm}[H]
\caption{The MIS algorithm of \cite{GhaffariGKMR18}.}
\label{alg:MISnew}
\begin{algorithmic}[1]
    \State{The leader $v^*$ generates a uniform random permutation $\pi\colon\interval{n}\mapsto\interval{n}$ and sends $\pi(i)$ to $v_i$.}
    \State{Each node $v_i$ broadcasts $\pi(i)$.
    }
    \State{$k\gets 0$}
    \While{The maximum active degree of active nodes is at least $\Delta'=\min\set{\Delta,\polylog{n}}$}
    \State{$M_0\leftarrow \emptyset$, $M_k \leftarrow M_{k-1}$ if $k\geq 1$}
    \State{$N_0\leftarrow \emptyset$, $N_k \leftarrow N_{k-1}$ if $k\geq 1$}
        \State{
        Every edge $\{v_i,v_{i'}\}$ with both endpoints  active and 
        $\pi(i)\leq \pi(i')\leq \frac{n}{\Delta^{\alpha^k}}$ is sent to $v^*$ by $v_i$. }
        \While{There exists a node $v_i \not\in M_k\cup N_k$ with $\pi(i)\leq \frac{n}{\Delta^{\alpha^k}}$}
            \State{Add $v_i$ with the smallest rank $\pi(i)$ among the undecided nodes to $M_k$.}
            \State{All the neighbors of $v_i$ that are known to $v^*$ are added to $N_k$.}
        \EndWhile
        \State{
        The leader $v^*$ informs the nodes in $M_k\setminus M_{k-1}$ that they are such.}
        \State{
        The nodes in $M_k\setminus M_{k-1}$ are added to $M$, and they inform their neighbors that they are  such and become inactive.}
        \State{
        The nodes in $N_G(M_k\setminus M_{k-1})$ inform their neighbors that they are such and become inactive.}
        \State{$k\gets k + 1$.}
    \EndWhile

     \For{
    $k$ from $0$ to $\bigO{\log\log\Delta'}$}
        \State{
        Each active node $v_i$ sends all adjacent edges from $H^{2^k}$ to each of its neighbors in $H^{2^k}$.}
    \EndFor
    
    \State{
    Each active node $v_i$  simulates $O(\log \Delta')$ rounds of the MIS protocol of \cite{Ghaffari2016} locally. The chosen nodes are added to $M$ and they become inactive along with their neighbors.
    }
    
    \State{
    The leader $v^*$ learns all remaining edges, locally computes an MIS over them and informs the nodes, which are then added to $M$.}
\end{algorithmic}
\end{algorithm}

In each iteration $k\geq 0$, we produce a set $M_k \subseteq V$, which is initially empty for $k=0$ and is initially $M_{k-1}$ for $k\geq 1$. The nodes in $M_k$ are afterwards added to the resulting MIS, $M$. We also use a set $N_k$ which is initially empty for $k=0$ and is initially $N_{k-1}$ for $k\geq 1$, of nodes that will not be in $M$. Initially, all nodes are \emph{active}. A node that is in $M_k\cup N_k$ is \emph{decided} and becomes \emph{inactive}, and otherwise it remains  \emph{active}.
 A constant $\alpha=3/4$ is set.

In each iteration $k$, all edges $\{v_i,v_{i'}\}$ where both endpoints are active and have ranks $\pi(i)\leq \pi({i'})\leq n/\Delta^{(\alpha^k)}$ are sent to the leader $v^*$ by $v_i$. The leader $v^*$ now applies greedy MIS steps, as follows. As long as there is an active node $v_i$ with $\pi(i)\leq n/\Delta^{(\alpha^k)}$, the node with the smallest rank is added to $M_k$  and all of its neighbors that are known to $v^*$ are added to $N_k$. After these greedy steps, the leader $v^*$ informs the nodes in $M_k$ that they are such. These nodes are added to $M$ and become \emph{inactive}, and they inform their neighbors, which join $N_k$ and become inactive as well.

The loop terminates when the maximum degree of active nodes is at most $\Delta'=\min\set{\Delta,\polylog{n}}$. To check that this condition is met, each node sends its degree to the leader and the leader broadcasts the decision. This requires $1$ round. We denote by $H=(V', E')$ the graph of maximum degree bounded by $\Delta'$ that is induced by the remaining active nodes.

\textbf{Small degrees (the graph \boldmath$H$):}
First, each active node generates $\bigO{\log^2{\Delta'}}$ random bits.
These random bits are from now sent along with the node's identifier whenever the latter is sent in a message. 
Then, each node of $H$ learns its $\bigO{\log{\Delta'}}$-hop neighborhood in $H$. To this end, we proceed in  $O(\log\log \Delta')$ iterations, where after iteration $k$, each node in $H$ knows its $2^k$-hop neighborhood in $H$. In iteration $k$, each node sends its edges in ${H}^{2^k}$ to its neighbors in ${H}^{2^k}$.
Notice that by induction over $k$, at the beginning of iteration $k$, each node knows its neighbors in ${H}^{2^k}$, and at the end of the iteration, it knows its neighbors in $H^{2^{k + 1}}$. 

After learning its $O(\log\Delta')$-hop neighborhood, each active node locally simulates $O(\log\Delta')$ rounds of the randomized MIS protocol of \cite{Ghaffari2016}. Each iteration of this protocol requires $\bigO{\log{\Delta'}}$ random bits by each node, which are the ones generated by the node at the beginning of this step. 
Each node chosen to the MIS is added to $M$ and becomes inactive along with its neighbors. Notice that all nodes compute the same MIS locally, because each node knows a sufficiently large neighborhood, including $O(\log^2\Delta')$ globally consistent random bits for each node in the neighborhood. 

\textbf{Wrapping-up part:} Finally, the leader $v^*$ learns the remaining graph induced by active nodes, and locally computes an MIS and informs the nodes, who are then added to $M$.
This finishes the description of the algorithm.

\paragraph{Message Complexity.}~

\textbf{Random ranking:} Since this part takes 2 rounds, it clearly sends at most $O(n^2)$ messages.

\textbf{Degree reduction (simulation greedy steps):}
Let $G_0$ be the subgraph induced by nodes with ranks $\pi(v_i)\leq \frac{n}{\Delta}$. Since the maximum degree in $G$ is $\Delta$, the number of edges in $G_0$ is bounded by $\frac{n}{\Delta}\cdot\Delta=n$. This implies that at most $O(n)$ messages are sent to the leader $v^*$ in the first iteration.

For $k\geq 1$, let $r_k=n/\Delta^{(\alpha^{k})}$, and let $G_k=(V_k, E_k)$ be the subgraph that is induced by nodes with ranks in the range $[r_{k-1}, r_k]$ that are still active after iteration $k-1$. In \cite[Theorem 1.1]{GhaffariGKMR18}, it is shown that $G_k$ has at most $O(n)$ edges, \whp, which implies that at most $O(n)$ messages are sent to the leader $v^*$ in iteration $k$. Informing the nodes in $M_k\setminus M_{k-1}$ that they should join $M_k$ requires at most $O(n)$ messages. Notice, that the leader does not inform nodes in $N_k\setminus N_{k-1}$, as they informed by their neighbors in $M_k\setminus M_{k - 1}$. Checking the loop condition required $\bigO{n}$ messages.
Since \cite[Lemma 3.1]{GhaffariGKMR18} implies that after $\bigO{\log\log\Delta}$ iterations of the loop, the degree in the graph induced by active nodes is at most $n\log{n} / (n / \Delta^{\alpha^{\bigO{\log\log\Delta}}})={\polylog{n}}=\Delta'$ and the loop terminates, this  gives a total of $O(n\log\log\Delta)$ of such messages, \whp. Over the entire execution of the protocol, each node $v_i$ is informed at most $\deg_G{(v_i)}$ times by one of its neighbors that such a neighbor enters the MIS or becomes inactive. Thus, over the entire course of the algorithm this requires $\bigO{n^2}$ messages.

\textbf{Small degrees:}
For $k=\bigO{\log\log\Delta'}$ the maximum degree in ${H}^{2^{k}}={H}^{{\polylog{\Delta'}}}$ is bounded by ${\Delta'}^{{\polylog{\Delta'}}}$. To send one identifier together with  $\bigO{\log^2\Delta'}$ random bits we need $1+O(\log^2\Delta'/\log n)$ $O(\log n)$-bit messages. Thus, for each $k$, each active node sends at most 
\begin{align*}
    O\big((1 + \log^2{\Delta'} / \log{n})\big){\Delta'}^{{\polylog{\Delta'}}}=2^{{\polylog{\Delta'}}}
\end{align*}
messages. 
For the entire $\bigO{\log\log\Delta'}$ rounds, each active node sends $\bigO{\log\log\Delta}'\cdot 2^{{\polylog{\Delta'}}}=2^{{\polylog{\Delta'}}}$ messages. As $\Delta'={\polylog{n}}$ holds, the number of messages sent by each active node to learn its $O(\log\Delta')$-hop neighborhood is $2^{{\polylog\log{n}}}=\bigO{n}$.
This implies $\bigO{n^2}$ messages in total. The simulation of \cite{Ghaffari2016} to decide whether to join $M$ is then done locally, without communication.

\textbf{Wrapping-up part:} 
In \cite[Lemma 2.11]{Ghaffari17}, it is shown that the graph induced by active nodes after learning $O(\log\Delta')$-hop neighborhoods in $H$ and simulating $O(\log \Delta')$ iterations of the MIS algorithm from \cite{Ghaffari2016} has at most $O(n)$ edges. Thus,  learning the remaining edges by the leader and informing nodes about the leader's decision requires $\bigO{n}$ messages. Notice that in the \CC model this would require some routing scheme. However, in the \CCL model this is part of the model definition.

Thus, overall the algorithm sends $\bigO{n^2}$ messages.
\end{proofof}

\subsection{Pointer Jumping}\label{subsec:pointerJumping}

In this subsection we address the pointer jumping problem, widely used in parallel and distributed data structures \cite{Hirschberg76}.

\begin{definition}[$P$-pointer jumping]
    In a \emph{$P$-pointer jumping problem}, each node $v_i$ is given a permutation $\pi_{i}\colon \interval{P}\mapsto\interval{P}$. A fixed node $v_{i'}$ is given a number $x\in\interval{P}$, The aim of the algorithm is for $v_{i'}$ to learn the composition of the permutations applied on $p$, i.e., $(\pi_{n - 1}\circ\pi_{n - 2}\circ\dots\circ\pi_0)(p)=\pi_{n - 1}(\pi_{n - 2}(\dots\pi_0(p)\dots))$
\end{definition}

In the following claim we show a simple deterministic $\bigO{\log{n}}$-round \CC protocol for solving $P$-pointer jumping with a complexity of $\bigO{P\cdot n}$ messages.

\begin{claim}[Pointer jumping]\label{claim:pointerJumping}
    For $P=\bigO{n}$, there is a deterministic $\bigO{P}$-memory efficient protocol in the \CCL model which solves the pointer jumping problem in $\bigO{\log{n}}$ rounds and $\bigO{P\cdot n}$ messages.
\end{claim}

\begin{algorithm}
\caption{Pointer jumping.}
\label{alg:pointerJumping}
\begin{algorithmic}
     \For{$k = 0$ to $\ceil{\log{n}}$}
        \For{$i\in\interval{n}$ in parallel}
            \If{$i$ has at least $k > 0$ trailing zeros in binary representation}
                \State{$v_i$ computes $\pi_{i}\circ\pi_{i + 1}\circ\dots\circ \pi_{\min\set{i + 2^{k}, n} - 1}$.}
            \EndIf
            \If{$i$ has exactly $k < \ceil{\log{n}}$ trailing zeros in binary representation}
             \State{$v_i$ sends $\pi_{i}\circ\pi_{i + 1}\circ\dots\circ \pi_{\min\set{i + 2^{k}, n} - 1}$ to $v_{i - 2^{k}}$.}
            \EndIf
        \EndFor
    \EndFor
    \State{$v_{i'}$ sends $p$ to $v_0$.
     }
    \State{$v_0$ replies $v_{i'}$ with $(\pi_{n - 1}\circ\pi_{n - 2}\circ\dots\circ\pi_0)(p)$.
       }
\end{algorithmic}
\end{algorithm}

\begin{proofof}[claim:pointerJumping]~

    \textbf{Algorithm and correctness.}
    The pseudo-code for the simple well known protocol for the pointer jumping problem is presented in \cref{alg:pointerJumping}. At a high level, first $v_0$ learns the composition of the permutations, then $v_{i'}$ sends the entry $p$ to $v_0$ and it responds with the final output. To learn the composition of the permutations we proceed in $\ceil{\log{n}} + 1$ iterations. On each iteration except the first, each node $v_i$ which receives a permutation from $v_{i + 2^{k - 1}}$, computes the composition of the  permutation it possesses and the received permutation, that is, it composes the permutation $\pi_i\circ\pi_{i + 1}\circ\dots\circ\pi_{i + 2^{k - 1} - 1}$ with the permutation $\pi_{i + 2^{k - 1}}\circ\dots\circ\pi_{i + 2^{k} - 1}$. Each node $v_i$ which has exactly $k$ trailing zeros in the identifier and currently knows the composition of $2^{k}$ permutations $\pi_i\circ\pi_{i + 1}\circ\dots\circ\pi_{\min\set{i + 2^{k}, n} - 1}$ sends it to the node $v_{i - 2^{k}}$. Each node sends and receives at most $P=\bigO{n}$ messages.
    Clearly, after $\ceil{\log{n}}$ iterations, node $v_0$ possesses the composition $\pi_0\circ\pi_1\circ\dots\circ\pi_{n - 1}$.
    
    \textbf{Memory-efficiency.}
    To compose the received permutation with the current permutation, we store both permutations and the output permutation in the local memory.
    Thus we require $\bigO{P\log{n}}$ bits of the local memory\footnote{In the \CC model this algorithm requires some routing scheme, but \cref{thr:optimalRouting} is not known to run in $\bigO{P\log{n}}$ bits of memory. However, our results apply in the potentially more powerful model of \CCL \cite{KuhnS20}, in which each node is allowed to send and receive $n$ messages in each round. Therefore, no additional memory overhead of routing algorithm is required. } . Given only the index of the round, it is possible for each node to deduce the number of messages each node sends to it.
    
    \textbf{Round complexity.}
    The algorithm finishes within $\bigO{\log{n}}$ iterations. In each iteration, each node sends and receives $P=\bigO{n}$ messages, thus each iteration completes in $\bigO{1}$ rounds.
    
    \textbf{Message complexity.}
    In $k$-th iteration of the algorithm, $\bigO{n / 2^k}$ nodes send $P$ messages each. Thus, the protocol uses $\sum_{k = 0}^{\ceil{\log{n}} - 1}{\bigO{n / 2^k}\cdot\bigO{P}}=\bigO{P\cdot n}$ messages.
\end{proofof}

Applying our deterministic scheduling algorithm and our random shuffling algorithm, we obtain the following theorem on the complexity of solving multiple instances of the pointer jumping problem.

\PJAmortized
\begin{proofof}[thr:PJAmortized]
    The first part of the theorem follows immediately from \cref{claim:pointerJumping,thr:optimalShedulingOfSmallJobs}.
     By \cref{thr:optimalShedulingOfSmallJobs}, running $t$ instances of the protocol from \cref{claim:pointerJumping} completes in $\bigO{t\cdot P\cdot n/n^2 + \ceil{P\cdot t / n}\cdot \log{n}}=\bigO{\ceil{P\cdot t / n}\cdot \log{n}}$ rounds. This gives the first claim.
    
    Since each node of the job consumes only $P\log{n}\leq n\log{n}$ bits of input and produces $\log{n}$ bits of output, it is $P$ I/O efficient. Thus, by \cref{thr:shuffle}, running $t$ instances of the protocol from \cref{claim:pointerJumping} completes in $\bigO{t + t\cdot P \cdot n / n^2 + \log{n}\cdot \log{n}}=\bigO{t + \log^2{n}}$ rounds, \whp, which gives the second claim.
\end{proofof}
The proposed simple $O(\log n)$ round pointer jumping protocol also serves as an example where scheduling jobs via the random-shuffling approach of \cref{thr:shuffle} is significantly better than the random-delay based approach of \cref{thr:delay}.
Since multiple nodes receive $\bigOmega{n\log{n}}$ messages in the execution of the protocol, if we apply the random-delay scheduling algorithm from \cref{thr:delay} we solve $t$ instances of the problem in  $\bigO{t\cdot n\cdot \log{n} + \log^2{n}}$ rounds, which is no better than sequentially running one instance after another.

\section{Discussion}
\label{sec:discussion}

Our results suggest that the amortized complexity, i.e., the runtime of solving many instances of a problem divided by the number of instances, is a valuable measure for the efficiency of protocols in the \CC model. Our interest in obtaining protocols with fast amortized complexities stems from the growing number of problems which admit $O(1)$-round \CC-protocols, e.g.,  \cite{CzumajDP20,Nowicki19,Ghaffari0T20}, whose amortized complexity could potentially be shown to go below constant, as well as from problems that are still not known to have a constant worst-case complexity. We now elaborate on this viewpoint.

We give MIS as an example of a problem which can be solved with a good amortized complexity. The best known protocol \cite{GhaffariGKMR18} requires $\bigO{\log\log\Delta}$ rounds. \Cref{thr:MISAmortized} shows that running $t=\poly{n}$ instances of MIS  completes in $\bigO{t + \log\log\Delta\log{n}}$ rounds. For $t=\bigOmega{\log\log\Delta\log{n}}$, the second part of the complexity ``amortizes out" and we obtain that we run $t$ instances of the MIS problem in $\bigO{t}$ rounds. Basically, we show that the amortized complexity of the MIS problem is $\bigO{1}$ rounds.

Note that the  amortized complexity should not be optimized isolated from other measures. For example,
consider the trivial $\bigO{n}$-round protocol for pointer jumping, in which in the $i$-th round, the $i$-th node applies its permutation to the \emph{current pointer} and sends the result to the next node. It requires only $\bigO{n}$ messages. Thus, it is trivial to run $t\leq n^2$ instances of this pointer jumping protocol in only $\bigO{n}$ rounds, leading to an amortized complexity of $\bigO{1 / n}=o(1)$. However, the 
\emph{latency}
of this algorithm is an unacceptable $\bigO{n}$ rounds.
Instead,  \cref{thr:PJAmortized} shows that the pointer jumping problem has an acceptable amortized complexity of $\bigO{1}$ rounds and a small latency of $\bigO{\log^2{n}}$ rounds. 

For certain protocols, \cref{thr:optimalShedulingOfSmallJobs} might even yield $o(1)$ amortized complexity.
For example, consider a job in which it is required to compute the $\sqrt{n}$-bin histogram of some given data. In the trivial $2$-round protocol, each node locally builds a histogram of its input and sends the number of elements in its $i$-th bin to $v_i$. For all $i\in[\sqrt{n}]$, node $v_i$ sums the received values and broadcasts the result. Clearly, such an algorithm is $\bigO{\sqrt{n}}$-memory efficient and uses $\bigO{n\sqrt{n}}$ messages. Our algorithm from \cref{thr:optimalShedulingOfSmallJobs} executes $t$ instances of this protocol in $\bigO{\ceil{t / \sqrt{n}}}$ rounds. Whenever $t=o(\sqrt{n})$, this gives an $o(1)$ amortized round complexity with constant latency.

The reader may notice that for some sets of jobs, it may be that some ad-hoc routing could be developed for efficient scheduling. We emphasize that, in contrast, the power of our algorithms is that they \emph{do not} require tailoring the protocols for the sake of scheduling them within a given set of jobs. This is pivotal for obtaining a general framework, because knowing in advance the setting in which a protocol would be executed is an unreasonable assumption that we do not wish to make.

\paragraph{Acknowledgements:} This project has received funding from the European Union’s Horizon 2020 research and innovation programme under grant agreement no. 755839-ERC-BANDWIDTH.

\bibliographystyle{alpha}

\bibliography{references}
\clearpage

\appendix

\end{document}